\documentclass[11pt,reqno]{amsart}
\usepackage{amsthm, amsmath,amsfonts,amssymb,euscript,hyperref,graphics,color}
\usepackage{graphicx}
\usepackage{import}
\usepackage{tikz}
\usepackage{latexsym}
\usepackage{cite}
\usepackage{verbatim}
\newcommand{\C}[1]{\textcolor{red}{#1}}

\newcommand{\nnb}{\nonumber}

\newcommand{\p}{\partial}

\newcommand{\D}{\mathcal{D}}

\newcommand{\di}{\mathrm{d}} 

\newcommand{\Eb}{\underline{E}}
\newcommand{\Lb}{\underline{L}}
\newcommand{\Cb}{\underline{C}}
\newcommand{\Cub}{\underline{C}_{\underline{u}}}
\newcommand{\ub}{\underline{u}}
\newcommand{\hb}{\underline{h}}

\newcommand{\ha}{\hat{w}}
\newcommand\hal{\hat{\alpha}}
\newcommand\halp{\hat{\alpha}^\prime}
\newcommand\alp{\alpha^\prime}
\newcommand\Php{\Phi^\prime}

\newcommand{\M}{\mathcal{M}}

\newcommand{\ove}{\overline}

\newcommand\al{\alpha}
\newcommand\be{\beta}
\newcommand\ga{\gamma}
\newcommand\Ga{\Gamma}
\newcommand\de{\delta}

\newcommand\ld{\lambda}
\newcommand\Ld{\Lambda}
\newcommand\si{\sigma}
\newcommand\Si{\Sigma}
\newcommand\ph{\phi}
\newcommand\Ph{\Phi}
\newcommand\ps{\psi}

\newcommand\Om{\Omega}
\renewcommand\th{\theta}
\newcommand{\vep}{\varepsilon}

 \def\<{\left\langle} \def\>{\right\rangle}
 \def\({\left(} \def\){\right)}

\newcommand{\bpf}{\begin{proof}}
 \newcommand{\epf}{\end{proof}}

\newcommand{\bg}{\begin}
 \newcommand{\ed}{\end}
 
\newcommand{\beq}{\begin{equation}}
\newcommand{\eeq}{\end{equation}}

\newcommand{\ali}[2]{
	\begin{align}\label{E:#1}
		#2
	\end{align}
}
\newcommand{\alis}[1]{
	\begin{align*}
		#1
	\end{align*}
}
\newcommand{\pM}[1]{
	\begin{pmatrix}
		#1
	\end{pmatrix}
}

\newtheorem{theorem}{Theorem}[section]
\newtheorem{lemma}[theorem]{Lemma}

\newtheorem{remark}[theorem]{Remark}

\numberwithin{equation}{section}

\allowdisplaybreaks[4]

\begin{document}
\title[Extension principle for SSEYM]{Extension principles for the Einstein Yang--Mills system}

\author[J. Li]{Junbin Li} 
\email{lijunbin@mail.sysu.edu.cn}
\address{Department of Mathematics, Sun Yat-sen University, Guangzhou 510275, China}

\author[J. Wang]{Jinhua Wang} 
\email{wangjinhua@xmu.edu.cn}
\address{School of Mathematical Sciences, Xiamen University, Xiamen 361005, China}

\begin{abstract}
We prove the local existence theorem and establish an extension principle for the spherically symmetric Einstein Yang--Mills system (SSEYM) with $H^1$ data. This in addition implies Cauchy stability for the system.

 In contrast to a massless scalar field, the purely magnetic Yang--Mills field in spherical symmetry satisfies a wave-type equation with a singular potential.  As a consequence, the proof of Christodoulou \cite{C-BV-93}, based on an $L^\infty-L^\infty$ estimate, fails in the Yang--Mills context. Instead, we employ an $L^2$-based method, which is valid for both massless and massive  scalar matter fields as well.
\end{abstract}
\maketitle

\section{Introduction}

 In  a series of works \cite{Chri-91, C-BV-93, Chri-94, Chri-99}, Christodoulou established a proof of the \emph{weak cosmic censorship conjecture}\footnote{Generic asymptotically flat initial data have a maximal future development possessing a complete future null infinity\cite{Chri-99-CQG}.} for the spherically symmetric Einstein massless scalar field system. This result was further demonstrated by Li-Liu \cite{Liu-Li-18, Li-Liu-JDG} and An \cite{An-aniso} under gravitational perturbations without symmetries. More recently, An-Tan \cite{An-Tan-WCC-EMKG} also verified the weak cosmic censorship conjecture for the Einstein charged massless scalar field system under spherical symmetry.

As a core step in the proof, Christodoulou \cite{Chri-99} introduced an instability mechanism for naked singularities within  the bounded variation (BV) topology. This framwork is well motivated, since both the local existence theory 
and the subsequent small-data global existence 
result for the spherically symmetric Einstein massless scalar field system had already been established in the BV topology \cite{C-BV-93}. In detailed proofs, one observes that the arguments for both (local or global) existence and instability rely crucially on a sharp criterion for $C^1$ extension \cite{C-BV-93, Chri-99}. 
This indicates that an extension principle in a suitable topology serves as the foundational element for addressing the weak cosmic censorship conjecture. 
Following the methodology of \cite{C-BV-93}, such extension criteria have also been established in the context of the Einstein charged massless scalar field and the Einstein massless Vlasov system under spherical symmetry\cite{An-Tan-WCC-EMKG, Moschidis-char-ib18}. 

We are now concerned with the Einstein Yang--Mills  system, which is of interest for several reasons. From a physical perspective, it triggers a rich variety of fascinated gravitational collapses, as evidenced by numerous numerical studies (see, for instance, \cite{BIZON-EYM-color, Choptuik-EYM-PRL, EYM-critical-2}). On the mathematical side, we demonstrate in this paper that the techniques required for the Einstein Yang–Mills system differ substantially from those employed in the study of gravity coupled with massless matter fields \cite{C-BV-93, Moschidis-char-ib18, An-Tan-WCC-EMKG}.

In this paper, we initiate the study of weak cosmic censorship conjecture for the Einstein Yang--Mills system in spherical symmetry. We will first consider the characteristic problem for the spherically symmetric Einstein Yang--Mills (SSEYM) system where the initial data are prescribed on a future light cone emanating from vertex at the center of symmetry. In this setting, we demonstrate the local existence theorem for data with a low-regularity norm ($H^1$ norm). We also deduce an extension principle for developments of $H^1$ data. Following this extension theorem, the Cauchy stability is justified for the Einstein Yang--Mills system under spherically symmetric perturbations, which are initially small with respect to the $H^1$ norms. 

As will be discussed later, the method used for the Einstein massless scalar field fails in the Einstein Yang--Mills scenario and hence an extension principle within the BV topology is not expected. The new $L^2$-based method we propose in this paper has a closer nature to the vacuum setting, and may be relevant for any future progress in vacuum. Indeed, our method is intimately tied to the H\"{o}lder topology that we are currently working in progress.

\subsection{Statement of the main results}
Let $\Ga$ be the set of fixed points of the symmetry group $SO(3)$, and $r$ be the area function of the group orbit, a 
$2$-sphere. Let $m$ be the Hawking mass contained within the sphere. 
We introduce the double null coordinates $(u, \, \ub)$, and denote $C_u$ and $\Cub$ the incoming and outgoing null cones.
For any $u_0 < \ub_\ast \in \mathbb{R}$, we define the domain of the quotient spacetime as
\begin{align}
 \D(u_0; \, \ub_\ast) :={}&  \{ (u, \, \ub) \, | \,u_0 \leq u < \ub_\ast, \,\, u \leq \ub < \ub_\ast \}, \label{def-D1} \\
  \D(u_0, \, u_1; \, \ub_\ast) := {}&  \{ (u, \ub) \, | \, u_0 \leq u \leq u_1, \,\, u \leq \ub \leq \ub_\ast \}. \label{def-D1-1}
\end{align}
As shown in their definitions, $\D(u_0; \, \ub_\ast) $ is the domain enclosed by the axis $\Ga$ and the hypersurfaces $C_{u_0}$, $\Cb_{\ub_\ast}$, while  $\D(u_0, \, u_1; \, \ub_\ast)$ is the domain bounded by  the axis $\Ga$ and the hypersurfaces $C_{u_0}$, $C_{u_1}$, $\Cb_{\ub_\ast}$.

We formulate the SSEYM system with the gauge group $SU(2)$ in Section \ref{sec-EYM}, where the \emph{spherically symmetric purely magnetic ansatz} for the Yang--Mills potential \eqref{SS-PM-ansatz} is assumed. 
 The characteristic problem of the SSEYM system involves prescribing a $C^0$ function $\p_r w|_{C_{u_0}}$ on an initial outgoing cone $C_{u_0}$ that emanates from vertex at $\Ga$. Here $w$ is related to the Yang--Mills potential \eqref{SS-PM-ansatz}. 
Throughout this paper, the $C^0$ Yang--Mills solution always refers to $\p_r w \in C^0$.
\begin{remark}\label{rk-data}
Without loss of generality, we assume $w|_{\Ga}=1$ and set 
\begin{equation}
\ha:= w-1. \label{def-ahat} 
\end{equation}
Then \[ \frac{\ha}{r}=\frac{1}{r}\int_0^r \p_r w \, \di r'|_{C_u}. \] This suggests that instead of specifying the initial datum $w|_{C_{u_0}}$, one may equivalently prescribe $\p_r w|_{C_{u_0}}$. 
\end{remark}

We define the $H^1$ norm for $\p_r w$
\beq\label{def-H1}
\|\p_r w\|_{H^1} :=\int_{C_{u} } \frac{1}{ r } \( \frac{\ha^2}{r^2} + (\p_r w)^2 + (r \p_r (\p_r w))^2 \)  \di r. 
\eeq
\bg{remark}\label{rk-equi-norm0}
It is remarkable that the norm \eqref{def-H1} is equivalent to 
\beq\label{def-H2-norm-1}
  \int_{ 0 }^r \frac{1}{r } \sum_{0\leq k\leq 2} \( (  r \p_r)^k   \frac{\ha}{r}  \)^2 \di r.
  \eeq
If we choose a new parameter $s := \ln r$,  \eqref{def-H2-norm-1} is alternatively written as 
\beq\label{def-H2-norm-2}
\int_{0}^{r_1} \frac{1}{r } \sum_{0 \leq k\leq 2} \( (  r \p_r)^k \frac{\ha}{r}  \)^2 \di r = \int_{-\infty}^{\ln r_1}   \sum_{0 \leq k\leq 2} \( \p_s^k  \frac{\ha}{r}   \)^2 \di s.
\eeq
The norm \eqref{def-H2-norm-2} is the standard $H^2$ norm for $\frac{\ha}{r}$. We abuse the notation a little bit and call \eqref{def-H2-norm-2} the $H^2$ norm of $\frac{\ha}{r}$, or rather $H^1$ norm of $\p_r w$. 
\ed{remark}
\bg{remark}\label{rk-moti-energy}
As a further remark, the standard energy for the Yang--Mills field does not align with the definition of energy given in \eqref{def-H1}. In fact, the conventional first-order energy for the Yang-Mills field takes the form $$\int_{C_{u} } \( \frac{\ha^2}{r^2} + (\p_r w)^2 \)  \di r,$$ which lacks the weight factor $\frac{1}{r }$. The motivation for \eqref{def-H1} stems from the right hand side of \eqref{eq-Lmu}: after integrating \eqref{eq-Lmu} in $\ub$, we aim to bound $\frac{m}{r}$ via the $H^1$ (in fact, $H^0$) norm of $\p_r w$.
\ed{remark}

We now state the $H^1$-extension theorem. A complete and detailed formulation is provided in Theorem \ref{extension thm-H1}.
\bg{theorem}\label{intro-extension thm-H1}
Suppose we have initial data $\p_r w|_{C_{u_0}}$ with bounded $H^1$ norm, 
and let $w$ be an $H^1$ solution (that is, $\p_r w \in H^1$) to the SSEYM system on the domain $\D(u_0; \, \ub_\ast)$.
There is a constant $\vep>0$ such that if \[ \lim_{u \rightarrow \ub_\ast} \sup_{\D(u; \, \ub_\ast)} \frac{m}{r} < \vep^2, \] the solution extends as an $H^1$ solution on the domain $\D( u_0; \, \ub_1)$ for some $\ub_1>\ub_\ast$.
\ed{theorem}

Following the extension theorem, we state a Cauchy stability result. A precise version is presented in Theorem \ref{stability-thm-1}.

\begin{theorem}\label{intro-stability-thm-1}
Suppose the data $\p_r w|_{C_{u_0}}$ admit a bounded $H^1$ norm that is small enough. Then, for any $\ub_0 \in (u_0, +\infty)$ and $ u_1\in (u_0, \, \ub_0]$, there exists a unique and global solution in $\D(u_0, \, u_1; \, \ub_0)$ whose $H^1$ norm remains small.
 \end{theorem}

\bg{remark}
In the subsection followed, we will sketch the main idea for the proofs.  We introduce the $L^2$-based method, tailored to the Yang--Mills field--or more precisely, to wave type equations with a positive, singular potential. This approach can also be adapted to both massless and massive scalar fields, ensuring that all the preceding statements (Theorems \ref{intro-extension thm-H1}--\ref{intro-stability-thm-1}) remain valid for the spherically symmetric Einstein (massive or massless) scalar field system. 

Since our proof is independ of $C^1$ bounds, but only depends on $H^1$ estimates, we effectively establish an extension principle in a weaker topology--the $H^1$ topology, compared to the $C^1$ framework used by Christodoulou \cite{C-BV-93}. Consequently, our results also require less regularity along the axis.
\ed{remark}

To facilitate the extension theorems, we establish a local existence theory for the characteristic problem of the SSEYM system in Section \ref{sec-local}. Additionally, for completeness, we present extension principle away from the axis in Appendix \ref{sec-extension-away-axis}. Thus, it confirms that the “first singularity” for the SSEYM system must orginate from the axis.

\subsection{Comments on the proof}

The purely magnetic Yang--Mills field obeys a wave-type equation with a singular potential proportional to $\frac{1}{r}$, which distinguishes it fundamentally from a standard massless scalar field. To highlight the difference, it is instructive to compare their formulations in Minkowski spacetime. The massless scalar field equation in Minkowski takes the form \[ \p_u \p_{\ub} (r \ph) =0, \] while the Yang--Mills equation is given by \[ \p_{u} \p_{\ub} w + \frac{w (w^2-1)}{r^2}=0. \] In terms of the variable  $\ha$ defined in \eqref{def-ahat}, the Yang--Mills equation can be reformulated as  \[ \p_u \p_{\ub} \ha +  \frac{2}{r}  \frac{\ha}{r} =  \frac{ \ha^2}{r^2} \(\ha+3\). \] 
This exhibits a nonlinear wave structure with an additional linear, singular term 
\[  \frac{2}{r}  \frac{\ha}{r},\] absent in the scalar wave case. As a result, the methods required to establish extension principles in the two settings necessarily differ. 

\subsubsection{The $H^1$ case}
To begin with, let us illustrate the proof for the massless scalar field \cite{C-BV-93}. Let $\al_s = \p_r (r \ph )$, where $\p_r$ is the null vector parallel to $\p_{\ub}$ and parametrized by $r$. The wave equation for $\ph$ leads to the following  equation for $\p_r \al_s$
\alis{
\p_u (\p_r \al_s) = {}& -2 \frac{\p_u r}{r} \frac{\frac{2m}{r}}{1-\frac{2m}{r}} \p_r \al_s + 3 \frac{\p_u r}{r^2} \frac{\frac{2m}{r}}{1-\frac{2m}{r}} r \p_{r} \ph  - \frac{\p_u r}{r^2} \frac{1}{1-\frac{2m}{r}} (r \p_{r} \ph )^3.
}
Remarkably, the right-hand side of this equation contains no linear terms in $\p_r \al_s$, $\al_s$, $r \p_r \ph$ and $\frac{2m}{r}$. It is exactly this structure that enables a usage of an $L^\infty-L^\infty$ estimate in the proof of the extension theorem \cite[Theorem 5.1]{C-BV-93}. 

To illustrate the key idea of this method, consider, without loss of generality, the term
 \[-2 \frac{\p_u r}{r} \frac{\frac{2m}{r}}{1-\frac{2m}{r}} \p_r \al_s. \] Its treatment exemplifies how nonlinear contributions are controlled within the $L^\infty$ framework.
In this quadratic term, the $\frac{2m}{r}$ offers a smallness factor $\vep$, rendering the entire term negligible. This allows an inequality of the form \[ \sup |\p_r \al_s| \leq \text{data} + \vep  \sup |\p_r \al_s| \] to yield a bound for $\p_r \al_s$. In principle, such $L^\infty-L^\infty$ estimates are effective for “generic” massless matter fields that share the characteristic absence of linear terms. Besides the massless scalar field, this $L^\infty-L^\infty$ approach has also been shown to apply to the charged massless scalar field \cite{An-Tan-WCC-EMKG} and the massless Vlasov system \cite{Moschidis-char-ib18}, to the best of our knowledge.

In contrast, for the Yang--Mills case, the presence of an additional potential leads to linear terms (linear in $\frac{\ha}{r}$ and $\al$) in the main equation for $\p_r \al$ \eqref{E:eq-pu-pr-al}, $\al := \p_r w$. Consequently, the  $L^\infty-L^\infty$ estimate strategy fails here, as these linear terms do not provide the required smallness.

Nevertheless, the positive sign of the potential in the Yang-Mills equation motivates the use of an $L^2$-based method. This approach allows us to handle quadratic terms arising from the linear contributions via integration by parts. After this manipulation, terms lacking inherent smallness give rise to quadratic expressions with definite signs. However, since our energy is weighted by $r^{-1}$ (and thus non-standard), these quadratic terms do not automatically exhibit favorable signs. In particular, the first-order energy identities \eqref{E:eid-E-1} and \eqref{E:eid-Eb-1} do not directly yield an energy bound. 

For this issue, we proceed to the second-order energy estimates and integrating by parts on the linear-derived terms. The second-order energy identity then produces quadratic terms with definite signs, which can potentially compensate for corresponding terms of opposite sign in the first-order identities (see \eqref{E:eid-E-2}). By strategically weighting these two energy identities, we ensure that all quadratic spacetime integrals become positive. This ultimately yields $H^1$ energy bounds (as detailed in Section \ref{subsec-energ-est-H2}), and pointwise estimates follow via a Sobolev-type inequality in the $H^1$ space. 

The method developed for the Yang-Mills case can also be applied to both massless and massive scalar fields, enabling the establishment of an extension principle in a low-regularity ($H^1$) setting. Furthermore, our approach allows less regularity near the axis: we only require $r \p_r \phi \sim 1$, while in Christodoulou \cite{C-BV-93}, the setting $r \p_r \phi \sim r$ near the axis is stronger. Additionally, the $H^1$ energy estimates play a crucial role in establishing the local existence result given in  Theorem \ref{thm-local-H2}.

\subsection{Outline of this paper}
This paper is organized as follows. In section \ref{sec-pre}, we formulate the SSEYM system. Section \ref{sec-H1} is devoted to establishing the extension theorem and Cauchy stability in $H^1$ space. The proof of local existence theorem is presented in Section \ref{sec-local}, while an extension principle away from the axis is provided in Appendix \ref{sec-extension-away-axis}.

\subsection{Acknowledgement} J.L. is supported by National Key R\&D Program of China (No. 2022YFA1005400) and NSFC (No. 12326602, 12141106). J.W. is supported by NSFC (No. 12271450).

\section{Preliminaries}\label{sec-pre}

\subsection{Spherically symmetric Einstein Yang--Mills equations}\label{sec-EYM}

In this section, we formulate the SSEYM system with  specific gauge assumption.

On a spherical symmetric spacetime $(\M, g)$, there exists an isometric action of the rotation group $SO(3)$, whose orbits are two-dimensional spheres. 
Define the axis $\Ga$ of $(\M, g)$ as the set of fixed points of the $SO(3)$ action.  We  introduce the double null coordinates $(u, \ub)$, and denote the level sets by $C_u$ and $\Cub$, representing the incoming and outgoing null cones respectively. The intersection $S_{\ub, u} = C_u \cap\Cub$ is a two-dimensional sphere. The area function $r$ is defined by \[\text{Area}(S_{\ub, u}) = 4 \pi r^2.\] The axis is characterized by \[ r=0 \quad \text{on} \, \, \Ga. \]  
The metric on $\M$ takes the form   \[ g= - 4 \Om^2 \di u \di \ub + r^2 \di \si_{S^2}, \] where $ \di \si_{S^2}$ is the standard metric on $S^2$ and the lapse function $\Om$ satisfies $-2 \Om^2 = g (\p_u, \p_{\ub})$.

We define \[ h=\Om^{-2} \p_{\ub} r, \quad \hb= \p_u r. \]
The Hawking mass contained within the $2$-sphere $S_{\ub, u}$ is defined via \[ m = \frac{r}{2} \( 1+ h \hb \). \] 
In the analysis approaching the axis, it is useful to define the mass ratio 
\beq\label{def-mu}
 \mu := \frac{2m}{r}. 
 \eeq  

In what follows, we turn to the Yang--Mills field with gauge group $SU(2)$. The group $SU(2)$ is the real Lie group of unitary matrices of determinant one, and its associated Lie algebra is $su(2)$, the anti-hermitian traceless $2\times 2$ matrices. Let $\tau_i$, $i \in \{ 1,2,3\}$,  be the following real basis of $su(2)$,  \[ \tau_1 = \frac{i}{2} \pM{0 & 1 \\ 1 & 0}, \, \, \tau_2 = \frac{1}{2} \pM{0 & -1 \\ 1 & 0}, \,\, \tau_3 = \frac{i}{2} \pM{1 & 0 \\ 0 & -1}, \]
with $[\tau_i, \, \tau_j] =  \epsilon_{ijk} \tau_k$, $ \epsilon_{ijk}$ being the Levi-Civita symbol, $\{i,j,k\} \subset \{1,2,3\}$.
We assume the \emph{spherically symmetric purely magnetic ansatz} \cite{BIZON-EYM-91, Choptuik-EYM-PRL} for the Yang--Mills potential, 
\beq\label{SS-PM-ansatz} 
A= w \tau_1 \di \th + \( \cot \th \tau_3 + w \tau_2 \) \sin \th \di \ph,  
\eeq
 where $(\th, \, \ph)$ are coordinates on $S^2$.
 Then the Yang--Mills field takes 
\alis{
F = {}& \di A + A \wedge A \\
={}& \di w \wedge \( \tau_1 \di \th + \sin \th \tau_2 \di \ph  \) - \(1-  w^2 \) \tau_3 \sin \th \di \th \wedge \di \ph.
}

As followed from the ansatz \eqref{SS-PM-ansatz}, $F_{u \ub} =0$, indicating that the electric charge vanishes;  The angular components are given by \[F_{AB} = \frac{w^2-1}{r^2} \tau_3 \vep_{AB},\] where capital Latin letters $\{A, B \}$ refer to indices on the 2-sphere $S^2$.  We define the magnetic charge within the sphere $S_{\ub, u}$ as 
\beq\label{def-charge}
Q: = w^2-1. 
\eeq
 The vacua of the Yang--Mills field correspond to the values $w=\pm 1$.

The Einstein Yang--Mills system takes the form 
\alis{
	R_{\al \be} - \frac{1}{2} R g_{\al \be} ={}& 2 T_{\al \be}, \\
	\hat{D}^\mu F_{\mu \nu} ={}& 0,
} 
where $\hat{D}_\mu = D_\mu + [A_\mu, \cdot \,]$ denotes the gauge covariant derivative, $[\, \cdot, \cdot \,]$ is the Lie bracket on $su(2)$, and the energy momentum tensor of Yang--Mills field is given by \[ T_{\al \be} = \<F_{\al}{}^{\nu}, \, F_{\be \nu} \> - \frac{1}{4} g_{\al \be} \<F_{\si \rho}, \, F^{\si \rho} \>. \] For more detailed descriptions for the Einstein Yang--Mills system, please refer to \cite{Choquet-Bruhat2009} and \cite{LW2021a, LW2021b}. 

We now present the spherical symmetric Einstein Yang--Mills system (SSEYM). With the ansatz \eqref{SS-PM-ansatz} and the double gauge, the SSEYM system reduces to the following equations for the variables $(r, \Om, w)$, 
\begin{subequations}
\begin{align}
\p_{\ub} h = {}& - 2 \Om^{-2} \frac{(\p_{\ub} w)^2}{r}, \label{eq-Lh} \\
\p_u (\Om^{-2} \hb) = {}& - 2 \Om^{-2} \frac{(\p_u w)^2}{r}, \label{eq-Lb-Omhb} \\
\p_{\ub} \hb = {}& - \Om^2  \frac{\mu}{r} + \Om^2 \frac{ Q^2}{r^3},  \label{eq-L-hb} \\
\p_u (\Om^2 h) = {}& - \Om^2  \frac{\mu}{r} + \Om^2 \frac{ Q^2}{r^3},   \label{eq-L-Omh} \\
\p_{\ub} \p_u \ln \Om ={}& \p_u \p_{\ub} \ln \Om =  \Om^2  \frac{\mu}{r^2} - 2 \Om^2 \frac{ Q^2}{r^4},  \label{eq-L-omb}
\end{align}
\end{subequations}
and the Yang--Mills equation 
\beq\label{YM}
\p_u \p_{\ub} w + \Om^2 \frac{Q}{r^2} w = 0.
\eeq

Moreover, we can derive from the above system the subsequent equations for the mass function 
\begin{subequations}
\begin{align}
\p_{\ub} m = {}& \Om^2 h \frac{Q^2}{2 r^2} -  \Om^{-2} \hb (\p_{\ub} w)^2, \label{eq-Lm} \\
\p_u m = {}& \hb \frac{Q^2}{2 r^2} - h (\p_u w)^2. \label{eq-Lbm}
\end{align}
\end{subequations}

A characteristic initial data set for the SSEYM system consists  of a triplet $(r, \Om, \p_r w)|_{C_{u_0}}$ prescribed on an outgoing cone $C_{u_0}$, as illustrated in Remark \ref{rk-data}. 
In addition to the initial condition, the SSEYM system is incorporated with the following boundary condition on the axis $\Ga$
\beq\label{bdy-condition}
r|_{\Ga}=0, \quad  m|_{\Ga} =0, \quad Q|_{\Ga} =0.
\eeq
The condition $Q|_{\Ga} =0$ is essential to ensure the regularity of the Yang--Mills equation \eqref{YM} on the axis.

We can reformulate the SSEYM system by eliminating $\Om$ and introducing the mass function $m$. The resulting equivalent system for the variables $(r, m, w)$ comprises the equations
\[ \eqref{eq-Lh},   \, \, \eqref{eq-Lb-Omhb}, \,\, \eqref{YM}, \,\,  \eqref{eq-Lm}, \,\, \eqref{eq-Lbm} \]
together with the boundary condition \eqref{bdy-condition}.

\subsection{Gauge choice}\label{sec-gauge}
We fix the coordinate $\ub$ on the initial outgoing cone $C_{u_0}$ by requiring that $\ub =  2 r + u_0$ on $C_{u_0}$, which suggests the gauge condition
\beq\label{gauge-ub}
\Om^2 h |_{C_{u_0}} = \frac{1}{2}. 
\eeq
Next, we choose the double null coordinates $(u, \, \ub)$ satisfying the additional gauge condition \[ u=\ub \quad \text{on} \,\, \Ga. \]  Since $r=0$ on $\Ga$, it follows that 
\beq\label{Om2h-hb-Ga} 
\Om^2 h = - \hb \quad \text{on} \,\,\Ga.
\eeq

\subsection{More equations}
For notational convenience, we define
\begin{align}
\al :={}& \Om^{-2} h^{-1} \p_{\ub} w, \label{def-al} \\
\alp:={}& r\Om^{-2} h^{-1} \p_{\ub} \al, \label{def-alp} \\
\Ph :={}& \frac{ Q }{ r}, \label{def-Ph} \\
\Php :={}& r \Om^{-2} h^{-1} \p_{\ub} \Ph. \label{def-Php}
\end{align}

The transport equations for $m$ \eqref{eq-Lm}--\eqref{eq-Lbm} help to deduce the subsequent equations for $\mu$,
\begin{subequations}
	\begin{align}
		\p_{\ub} \mu = {}& - \Om^2 h \frac{\mu}{r} + \Om^2 h \frac{\Ph^2}{ r} + 2 \( 1-\mu \)  \Om^{2} h  \frac{\al^2}{r}, \label{eq-Lmu} \\
		\p_u \mu = {}&  - \hb \frac{\mu}{r} + \hb \frac{\Ph^2}{ r} - 2 h \frac{(\p_u w)^2}{r}. \label{eq-Lbmu}
	\end{align}
\end{subequations}

Due to the Yang--Mills equation \eqref{YM}, we have
\beq\label{eq-pu-al}
\p_u \al = - h^{-1} w \frac{Q}{r^2}+ h^{-1} \( \frac{\mu}{r}- \frac{Q^2}{r^3} \) \al.
\eeq
Along with \eqref{eq-pu-al}, we need a higher order equation

\ali{eq-pu-pr-al}{ \p_u ( \p_r \al) 
={}&\underbrace{ - 2w^2 \frac{h^{-1}}{r^2}  \al + 2 w \frac{ h^{-1}}{r^2} \frac{Q}{r}}_{\text{Linear terms}} \nnb \\
& + 2 \frac{h^{-1}}{r} \( \mu - \Ph^2 \)  \p_r \al   - \frac{h^{-1}}{r} \Ph \al - 6 w \frac{h^{-1}}{r^2}   \Ph \al^2  \nnb\\
& +2 \frac{h^{-1}}{r^2} \(  2 \Ph^2 -  \mu  \) \al   + 2 \frac{h^{-1}}{r^2} \( 1 - \Ph^2 \)  \al^3. 
} 
 We always denote \[ \p_r := \Om^{-2} h^{-1} \p_{\ub}\] the outgoing null vector parametrized by $r$.

\section{Extension theorem in $H^1$ space}\label{sec-H1}
We aim to establish an extension theorem for solutions to the SSEYM system in the class of $\( r, \, m, \, \al \) \in C^1 \times C^0 \times H^1$. Recall \eqref{def-H1} for the definition of   $H^1$ norm.

For any $u_0 < \ub_\ast \in \mathbb{R}$,  the domain $ \D(u_0; \, \ub_\ast)$ bounded by the axis $\Ga$, $C_{u_0}$ and $\Cb_{\ub_\ast}$, is defined in \eqref{def-D1}. 

\bg{theorem}\label{extension thm-H1}
Given initial data $\al|_{C_{u_0}}$, with finite initial energy 
\[ \int_{C_{u_0} \cap \{ u_0 \leq \ub \leq \ub_0 \} } \frac{1}{ r} \( \al^2 + (\alp)^2 + r^{-2} \ha^2 \)  \di r, \quad \text{for any} \,\,\, \ub_0>u_0,  \] suppose we have on the domain $\D(u_0; \, \ub_\ast)$ a solution to the SSEYM system $(r, \, m, \, \al) \in C^1 \times C^0 \times H^1$.
There is a constant $\vep>0$ such that if \[ \lim_{u \rightarrow \ub_\ast} \sup_{\D(u; \, \ub_\ast)} \mu < \vep^2, \] then the solution extends to $\D( u_0; \, \ub_1)$, for some $\ub_1>\ub_\ast$, with the same regularity $C^1 \times C^0 \times H^1$. 

Moreover, the solution obeys the estimates on $\D( u_0; \, \ub_1)$
\begin{align}
|r^{-1} \ha|^2 \lesssim \vep^2, \quad \text{and} \quad |\al|^2 \lesssim  \vep, \label{Pointwise-bound} 
\end{align}
and the energies remain bounded
\ali{Energy-bound}{
&\int_{C_u \cap \D( u_0; \, \ub_1)} \frac{1}{r} \( \Ph^2 + \al^2 + (\alp)^2  \) \di r \nnb \\
&+ \int_{\Cub \cap \D( u_0; \, \ub_1)}  \frac{h^{-1}}{r} \( \Ph^2 + \al^2 + h^2 (\p_u w)^2 \)  \di u <C.
}

\ed{theorem}

In the rest of this section, we will prove Theorem \ref{extension thm-H1}.

\subsection{Conventions}
Let us introduce some conventions for this section. For any $u_0 < u_1 \leq \ub_\ast$, we recall the definition of the domain $\D(u_0, \, u_1; \, \ub_\ast)$  from \eqref{def-D1-1}, which is bounded by $\Ga$ and the hypersurfaces $C_{u_0}$, $C_{u_1}$, $\Cb_{\ub_\ast}$.
We make the following simplification 
\beq\label{def-domain2}
 \D :=  \D( u_0, \, u_1; \, \ub_\ast + \zeta ), \quad \text{with some constant} \,\, \zeta>0,
\eeq  
and denote the truncated null cones as \[ C_u^{[\ub_0, \,\ub_1]} :=C_u \cap \{  \ub_0 \leq \ub \leq \ub_1 \}, \quad \Cub^{[u_0,\, u_1]} :=\Cub \cap \{  u_0 \leq u \leq  u_1 \}.  \] Furthermore, we abbreviate the outgoing cone segment emanating from the vertex $C_u^{[u,\, \ub]}$ as $C_u$, and $\Cub^{[u_0, \, u]}$ as $\Cub$, when the interval of integration is clear in the context. 

We define the energy norms,
\ali{def-E-Eb}{
E_0(u, \, \ub) :={} &\int_{C_u } \frac{\Om^2 h}{r} \( r^{-2} \ha^2 + \al^2 \) \di \ub^\prime, \nnb \\
E_1(u, \, \ub) :={} &  E_0(u, \, \ub) + \int_{C_u } \frac{\Om^2 h}{r}  (\alp)^2 \di \ub^\prime, \nnb \\
\Eb_0(u, \, \ub):= {}& \int_{ \Cub }  \frac{h^{-1}}{r} \( r^{-2} \ha^2 + h^2 (\p_u w)^2 \)  \di u', \nnb \\
\Eb_0^\prime(u, \, \ub):= {}&  \Eb_0(u, \, \ub)+  \int_{ \Cub }  \frac{h^{-1}}{r} \al^2 \di u'.
}
If we use $r$ as the parameter on $C_u$, $\Cub$ respectively, then the energies in \eqref{E:def-E-Eb} are rewritten as
\alis{
E_0 ={}& \int_{0}^r \frac{ r^{-2} \ha^2 + \al^2 }{r} \di r|_{C_u}, \\
E_1 ={}& \int_{0}^r \frac{(\alp)^2 + \al^2 +  r^{-2} \ha^2 }{r} \di r|_{C_u}, \\
\Eb_0 ={}& \int_{r}^{r_0} \frac{1}{1-\mu} \frac{ r^{-2} \ha^2 + h^2 (\p_u w)^2}{r} \di r|_{\Cub}, \\
\Eb_0^\prime ={}&  \Eb_0 + \int_{r}^{r_0} \frac{1}{1-\mu} \frac{\al^2}{r} \di r|_{\Cub},
}
where $r_0 := r(u_0, \ub)$.

\subsection{Energy estimates theorem}\label{sec-energy-estimate-H2}
The proof of Theorem \ref{extension thm-H1} follows from the energy estimates which we state as below.

\bg{lemma}\label{lem-extension}
Suppose $(r, \, m, \, \al)$ is any solution to the SSEYM system in $\D$ for any $u_0 < u_1 \leq \ub_\ast$ and some constant $0<\zeta<+\infty$. Given initial data with the bounded  energy \[ E_1 ( u_0, \, \ub_\ast ) \leq I_1, \quad \text{for some} \, I_1 \in \mathbb{R}_{>0}. \]  The subscript $1$ in $I_1$ denotes the number of derivatives used in the energy norms. There is an $\vep>0$ such that if \[ \mu \leq \vep^2, \quad \text{in} \,\,  \D, \]  the  solution $(r, m, \al) \in C^1 \times C^0 \times  H^1$ in $\D$. Moreover, it holds that 
\beq\label{lem-pointw-1} 
|r^{-1} \ha|^2 \leq C( I_1) \vep^2, \quad  |\al|^2 \leq C( I_1) \vep, 
\eeq
and the energies are uniformly bounded
\ali{lem-energy-bound}{
E_1 (u, \,\ub) + \Eb_0^\prime (u,\, \ub) \leq C( I_1), \quad \text{where} \,\,\,  u \in [u_0,\, u_1],  \,\, \ub \in [u, \, \ub_\ast + \zeta ]
} for some constant $C(I_1)$ depending only on $I_1$.

\ed{lemma}

\bg{remark}
The energy arguments leading to Lemma \ref{lem-extension} are closed in the $H^1$ space. This fact enables us to prove a local well-posedness theorem with $H^1$ data, as detailed in Section \ref{sec-local}.
\ed{remark}

\subsection{Preliminary estimates}\label{sec-pre-estimate-H2}
                                                                
As a preliminary step, we have 
\beq\label{sign-h-hb}
  \hb<0, \quad h>0 \quad \text{in} \,\, \D.
\eeq
The fact $\hb <0$ follows from the monotonicity of $\Om^{-2} \hb$ in $u$ (see \eqref{eq-Lb-Omhb}) and the initial requirement $\Om^{-2} \hb|_{u_0} <0$, and moreover, $h>0$ is due to the smallness of $\mu$ which indicates $1-\mu = - \hb h>0$.

\subsection{Continuity argument}
We make the following bootstrap assumption. Let $\Ld$ be a large constant that will be determined (depending on the initial energy) later. We assume
\beq\label{BT-H1}
 |r^{-1} \ha|^2 + |\al|^2   \leq  \vep \Ld, \quad \text{in} \,\, \D.
\eeq
and will show that the same inequality with $\Ld$ being replaced by $\frac{\Ld}{2}$ holds.

Since $r$ is finite in $\D$, it is straightforward to know from \eqref{BT-H1} that
\beq\label{esti-w-1}
|\Ph|^2 \lesssim \vep \Ld,\,\, \text{and} \,\, |w^2-1|^2 \lesssim \vep \Ld, \quad \text{in} \,\, \D.
\eeq

\subsection{Main energy estimates}\label{subsec-energ-est-H2}
We will use the bootstrap assumption \eqref{BT-H1} to carry out the energy estimates in details.

Let us begin with the first order energy identities.
We multiply $2 \p_{\ub} w$ on the Yang--Mills equation \eqref{YM} to obtain the identity
\beq\label{eq-pu-lw2}
\p_u (\p_{\ub} w)^2 + \frac{\Om^2}{2 r^2} \p_{\ub} Q^2=0.
\eeq
We further multiply $\Om^{-2} h^{-1} r^{-1}$ on \eqref{eq-pu-lw2}, then
\alis{
	& \p_u \( r^{-1} \Om^{-2} h^{-1} (\p_{\ub} w)^2 \) - \p_u r^{-1} \Om^{-2} h^{-1} (\p_{\ub} w)^2 \\
	&\qquad  - \p_u (\Om^{-2} h^{-1}) \frac{(\p_{\ub} w)^2}{r } + \p_{\ub} \( \frac{h^{-1}}{2r^{3}} Q^2 \) - \p_{\ub} \( \frac{h^{-1}}{2r^{3}}  \) Q^2=0.
}       
Using the equations \eqref{eq-Lh} and \eqref{eq-L-Omh}, and integrating over the domain $\D$,
we arrive at the following energy identity
\ali{eid-E-1}{
	&\int_{C_u} \Om^2 h \frac{\al^2}{r } \di \ub + \int_{\Cub} h^{-1} \frac{ \Ph^2 }{2 r } \di u \nnb \\
	&+ \iint_{\D} \( \frac{3}{2}  \frac{ \Ph^2 }{r^{2}} -  \frac{\al^2}{r^{2}} \) \Om^2 \di u \di \ub \nnb \\
	={}& \int_{C_{u_0}} \Om^2 h \frac{\al^2}{r} \di \ub.
} 
Note that, among the spacetime integrals on the left-hand side of \eqref{E:eid-E-1}, there appears a negative term, which obstructs the derivation of  energy bounds at this stage. 

On the other hand,  multiplying $2 \p_u w$ on the Yang--Mills equation \eqref{YM} yields
\beq\label{eq-pub-Lbw2}
\p_{\ub} (\p_u w)^2 + \frac{\Om^2}{2 r^2} \p_u Q^2=0.
\eeq
Multiplying by $h r^{-1}$ and using equations \eqref{eq-Lh}, \eqref{eq-L-Omh}, we transform \eqref{eq-pub-Lbw2} into the following identity, 
\alis{
	& \p_{\ub} \( h  \frac{ (\p_u w)^2 }{r} \) + \Om^2 h^2 \frac{(\p_u w)^2}{r^{2}} + 2 \Om^{-2} ( \p_{\ub} w)^2 \frac{ (\p_u w)^2 }{r^{2}}   \\
	& \quad + \p_u \(\frac{\Om^2 h}{2} \frac{ Q^2}{r^{3}} \) + \frac{3}{2} \Om^2 \hb h \frac{Q^2}{r^{4}} + \frac{\Om^2}{2} \( \frac{\mu}{r} - \frac{Q^2}{r^3} \)  \frac{Q^2}{r^{3}}=0.
}
We then obtain an energy identity, after integrating over the domain $\D$,
\ali{eid-Eb-1}{
	&\int_{\Cub}  h \frac{(\p_u w)^2}{r} \di u + \int_{C_u}  \Om^2 h \frac{ \Ph^2 }{2 r} \di \ub \nnb \\
	&+ \iint_{\D} \( 1 + 2  \al^2  \) \Om^2 h^2 \frac{(\p_u w)^2}{r^{2}} \di u \di \ub \nnb \\
	& +  \iint_{\D} \(  - \frac{3}{2} + 2 \mu  \) \Om^2 \frac{ \Ph^2 }{r^{2}} \di u \di \ub \nnb \\
	={}& \int_{C_{u_0}} \Om^2 h \frac{ \Ph^2 }{2 r} \di \ub + \iint_{\D}  \Om^2  \Ph^2 \frac{ \Ph^2 }{2r^{2}} \di u \di \ub.
}

Analogous to \eqref{E:eid-E-1}, the identity above contains a nonpositive spacetime integral on the left-hand side, which prevents it from yielding an energy bound. What is more, cooperating the  energy identity \eqref{E:eid-E-1} with \eqref{E:eid-Eb-1} does not give any bound for the first order energies neither. 

To address this, we now derive the second order energy identities, which are designed to absorb the terms with unfavorable signs in \eqref{E:eid-E-1} and \eqref{E:eid-Eb-1}.

By virtue of \eqref{E:eq-pu-pr-al}, we have
\ali{eq-pu-alp}{
	\p_u \alp ={}& \frac{\hb}{r} \alp   - 2  w^2 \frac{h^{-1}}{r} \al  + 2 w \frac{ h^{-1}}{r} \frac{Q}{r} \nnb \\
	& + 2 \frac{ h^{-1}}{r} \( \mu - \Ph^2 \) \alp   -  h^{-1} \Ph \al  - 6 w \frac{h^{-1}}{r} \Ph \al^2  \nnb  \\
	& + 2 \frac{h^{-1}}{r} \(2 \Ph^2 - \mu \) \al + 2 \frac{ h^{-1}}{r} \(1  - \Ph^2 \)   \al^3,
}  
where we mark the linear terms on the first line as below
\alis{
	\ell_1= \frac{\hb}{r} \alp, \quad \ell_2 =  - 2 w^2 \frac{h^{-1}}{r} \al, \quad \ell_3 =  2 w \frac{ h^{-1}}{r} \frac{Q}{r}.
}

Before proceeding to the detailed second order energy estimates, we outline the main idea of the proof.
To obtain an energy identity from \eqref{E:eq-pu-alp}, we multiply the equation by \[ r^{-1} \cdot \Om^{2} h \alp = \p_{\ub} \al \] and integrate over $\iint_{\D} \di u \di \ub$. Regarding to the first terms on the right-hand side of the first line,  
\beq\label{eid-2nd-main}
\iint_{\D}  \Om^{2} h \frac{\alp}{r}\cdot \ell_1 \di u \di \ub= - \iint_{\D} \( 1-\mu \) \frac{(\alp)^2}{r^{2}} \Om^2 \di u \di \ub
\eeq 
admits a favourable sign and can be moved to the left-hand side. In contrast, the remaining terms $\iint_{\D} r^{-1} \cdot \Om^{2} h \alp \cdot \ell_i \, \di u \di \ub$, $i=2, \, 3$, which are now quadratic, lack definite signs.  A straightforward estimate such as 
\alis{
	& \iint_{\D}  \Om^{2} h \frac{\alp}{r} \cdot \ell_i \, \di u \di \ub \\
	\lesssim & \( \iint_{\D} \Om^{2} \frac{(\alp)^2}{r^{2}} \di u \di \ub \)^{\frac{1}{2}} \cdot \( \iint_{\D}  \Om^{2} \( \frac{\al^2}{r^{2}} + \frac{\Ph^2}{r^{2}} \) \di u \di \ub \)^{\frac{1}{2}}
}
is not allowed at this stage, as we do not yet have improved bounds for the spacetime integrals on the right-hand side.
The difficulty posed by these quadratic terms is resolved via applying integration by parts. This allows us to ultimately obtain spacetime integrals with favorable signs, such as $\iint_{\D}  \Om^{2} \frac{\al^2}{r^{2}}   \di u \di \ub$, $\iint_{\D}  \Om^{2} \frac{\Ph^2}{r^{2}} \di u \di \ub$. 

Among all the terms on the right-hand side, the nonlinear terms in the second and third lines can, in principle, be estimated by means of (after multiplication by $r^{-1} \cdot \Om^{2} h \alp $) 
\beq\label{eid-2nd-cubic}
\lesssim \vep \Ld \iint_{\D} \( \frac{\Ph^2}{r^{2}} + \frac{\al^2}{r^{2}} + \frac{(\alp)^2}{r^{2}} \) \Om^2 \di u \di \ub.
\eeq
Actually, the term \eqref{eid-2nd-cubic} is expected to be absorbed, since during the treatment of the linear terms $\ell_i$, we have already managed to generate positive bulk terms of the form $\iint_{\D} \( \frac{\Ph^2}{r^{2}} + \frac{\al^2}{r^{2}} + \frac{(\alp)^2}{r^{2}} \) \Om^2 \di u \di \ub$ on the left-hand side of the energy inequality via integration by parts. 

We now go on with the detailed proof. First of all, we analyze the left-hand side of \eqref{E:eq-pu-alp}. After multiplied $\Om^2 h \frac{\alp}{r}$, it becomes,
\alis{
	&  \p_u \alp \cdot \Om^2 h \frac{\alp}{r } \\
	={}&  \p_u \( \Om^2 h \frac{(\alp)^2}{2 r } \) - \p_u \( \Om^2 h \) \frac{(\alp)^2}{2 r }  - \frac{1}{2} \p_u \( \frac{1}{r } \) \Om^2 h (\alp)^2 \\
	={}&  \p_u \( \Om^2 h \frac{(\alp)^2}{2 r } \) + \Om^2  \( \frac{\mu}{r} - \frac{Q^2}{r^3} \) \frac{(\alp)^2}{2 r }  +  \frac{1}{2}   \hb h \Om^2 \frac{(\alp)^2}{ r^{2}}.
} 
We observe that the last term \[ \frac{1}{2}   \hb h \Om^2 \frac{(\alp)^2}{ r^{2}} = - \frac{1}{2} (1-\mu)\Om^2 \frac{(\alp)^2}{ r^{2}} \] on the left-hand side of the energy identity carries an unfavorable sign. Nevertheless, this term can be absorbed by the positive contribution from \eqref{eid-2nd-main}, and ultimately yields a net positive term \[   \iint_{\D} \frac{1}{2} \( 1-\mu \) \frac{(\alp)^2}{r^{2}} \Om^2 \di u \di \ub \]  on the left-hand side of the energy identity.

To proceed, we focus on the two terms $\iint_{\D} r^{-1}  \Om^{2} h \alp \cdot \ell_i \, \di u \di \ub$, $i=2, \, 3$. When $i=2$, it holds that
\alis{
	& r^{-1}  \Om^{2} h \alp \cdot \ell_2 = - w^2 h^{-1} r^{-1} \p_{\ub} \al^2 \\
	={}& - \p_{\ub} \( w^2 h^{-1} r^{-1} \al^2 \) + \p_{\ub} w^2  h^{-1} r^{-1} \al^2 \\
	&+ w^2 r^{-1} \al^2 \p_{\ub} h^{-1} + w^2 \p_{\ub} r^{-1} \al^2 h^{-1} \\
	={}&  - \p_{\ub} \( w^2 h^{-1} r^{-1} \al^2 \) + 2 \Om^2 w r^{-1} \al^3 \\
	&+2 w^2  \Om^{2}  r^{-2} \al^4 -  w^2 \Om^{2} r^{-2} \al^2,
}  where we have made use of the equation \eqref{eq-Lh} in the last identity.     
In particular, we note that the last term $- w^2 \Om^{2} r^{-2} \al^2$ on the right-hand side the energy identity has a favourable sign. When $i=3$, we have
\alis{
	& r^{-1}  \Om^{2} h \alp \cdot \ell_3 = 2 h^{-1} r^{-2} (w^2-1) w \cdot \p_{\ub } \al \\
	={}& \p_{\ub} \( 2 r^{-2} h^{-1} (w^2-1) w \al \) - \p_{\ub} \( 2 r^{-2} h^{-1} (w^2-1) \) w \al \\
	&- \p_{\ub} w \cdot \( 2 r^{-2} h^{-1} (w^2-1)  \al \) \\
	={}&  \p_{\ub} \( 2 r^{-2} h^{-1} (w^2-1) w \al \)  + 4   \Om^2 r^{-3} (w^2-1) w \al \\
	&- 4 h^{-2} \Om^{-2} \frac{(\p_{\ub} w)^2}{r^{3}} (w^2-1) w \al - 4 \Om^2 \frac{\al^2}{r^{2}} - 6 (w^2-1) \Om^2 \frac{\al^2}{r^{2}}.
}
Here the quadratic term $- 4 \Om^2 \frac{\al^2}{r^{2}}$ admits a favourable sign, whereas $4\Om^2 \frac{w^2-1}{r^{3}} w \al$ lacks a definite sign. Nevertheless, we can perform the same technique to this quadratic term to proceed
\alis{
	& 4\Om^2 \frac{w^2-1}{r^{3}} w \al=  2h^{-1} \frac{w^2-1}{r^{3}} \p_{\ub} (w^2-1) \\
	={}& \p_{\ub} \( h^{-1} \frac{(w^2-1)^2}{r^{3}} \) -  \p_{\ub} h^{-1} \frac{(w^2-1)^2}{r^{3}} + 3 \Om^2\frac{ (w^2-1)^2}{ r^{4} }.
}  
We then combine these calculations to obtain
\alis{
	& r^{-1}  \Om^{2} h \alp \cdot \ell_3 \\
	={}&  \p_{\ub} \( 2 w h^{-1} \frac{\Ph  \al}{r} \)  + \p_{\ub} \( h^{-1} \frac{\Ph^2}{r} \)   \\
	& - 4 \Om^2 \frac{ \al^2 }{r^{2}} + 3  \Om^2 \frac{\Ph^2}{r^{2}} - 2 \Om^{2} \frac{ \al^2}{r^{2}} \Ph^2 - 4 \Om^{2} \frac{\Ph}{r^{2}}  w \al^3 - 6 \Om^2 \Ph \frac{\al^2}{r}.
}  
In the third line, the quadratic term $- 4 \Om^2  \frac{ \al^2 }{r^{2}}$ contributes positively to the energy identity, while $3 \Om^2 \frac{\Ph^2}{r^{2}}$ has an unfavorable sign. Nevertheless, noting the presence of the positive integral $\iint_{\D} \frac{\Ph^2}{r^{2}} \di u \di \ub$ on the left-hand side of the first order energy identity \eqref{E:eid-E-1}, the term $3 \Om^2 \frac{\Ph^2}{r^{2}}$ can be absorbed when \eqref{E:eid-E-1} is taken into accounts.

Integrating over $\iint_{\D} \di u \di \ub$ leads to the energy identity
\ali{eid-E-2}{
	& \int_{C_u}  \frac{(\alp)^2}{2 r}  \Om^2 h \di \ub + \int_{\Cub} \frac{1}{r} \( \(   w \al - \Ph \)^2 - 2 \Ph^2  \) h^{-1} \di u \nnb \\
	& + \iint_{\D}  \( \frac{1}{2} + \frac{3\Ph^2}{2}  \) \frac{(\alp)^2}{ r^{2}} \Om^2 \di u \di \ub \nnb \\
	& + \iint_{\D}  \( 5 + 2 \Ph^2  \) \frac{\al^2}{r^{2}} \Om^2 \di u \di \ub  - \iint_{\D} 3   \frac{\Ph^2}{r^{2}} \Om^2 \di u \di \ub \nnb \\
	={}& \int_{C_{u_0}}  \frac{(\alp)^2}{2 r}  \Om^2 h \di \ub +\iint_{\D}   2 \mu \frac{(\alp)^2}{r^{2}}   \Om^2 \di u \di \ub \nnb \\
	& + \iint_{\D}  \( 2( 1 - \Ph^2) \al^2  +  4 \Ph^2 - 2\mu  -  6 w \Ph \al  -  r \Ph \) \frac{\al \alp}{r^{2}}   \Om^2 \di u \di \ub \nnb \\
	& +\iint_{\D} \(  2 \(  w  r \al +  w^2 \al^2  \) - 4w \Ph  \al - 7 r \Ph  \)  \frac{ \al^2}{r^{2}} \Om^2 \di u \di \ub.
}
Now in view of the identities \eqref{E:eid-E-1}, \eqref{E:eid-Eb-1} and \eqref{E:eid-E-2}, we aim to eliminate the negative term $- \iint_{\D} 3  \Om^2 \frac{\Ph^2}{r^{2}} \di u \di \ub$ on the left-hand side of \eqref{E:eid-E-2}, and meanwhile obtain a coercive energy on $\Cub$. To achieve this, we form the linear combination \[  9 \eqref{E:eid-E-1} + 2 \eqref{E:eid-E-2}  + 2 \eqref{E:eid-Eb-1}, \] which leads to
\ali{eid-sum2}{
	&\int_{C_u} \frac{ \Om^2 h }{r} \( 9 \al^2 + (\alp)^2 + \Ph^2 \) \di \ub \nnb\\
	& + \int_{\Cub}  \frac{h^{-1}}{r } \( 2 \(   w \al - \Ph \)^2 + \frac{ \Ph^2 }{2} + 2 h^2 (\p_u w)^2 \)  \di u \nnb \\
	&+ \iint_{\D}  \(     1 +  3 \Ph^2  \)  \frac{(\alp)^2 }{r^{2}} \Om^2 \di u \di \ub  + \iint_{\D} \( 1 + 4  \Ph^2    \)  \frac{\al^2 }{r^{2}} \Om^2 \di u \di \ub  \nnb \\
	&+ \iint_{\D}   \( \frac{9}{2} + 4\mu   \) \frac{\Ph^2 }{r^{2}} \Om^2 \di u \di \ub   + \iint_{\D} \(   2 +  4  \al^2  \)   \frac{h^2  (\p_u w)^2 }{r^{2}} \Om^2 \di u \di \ub  \nnb \\
	= {}& \int_{C_{u_0}} \frac{ \Om^2 h }{r} \( 9  \al^2 + (\alp)^2 + \Ph^2 \) \di \ub + \iint_{\D}   \Ph^{2} \frac{\Ph^2}{r^{2}}  \Om^2 \di u \di \ub \nnb \\
	& +  \iint_{\D} 2 \( 2( 1 - \Ph^2) \al^2  +  4 \Ph^2 - 2\mu  -  6 w \Ph \al  -  r \Ph \) \frac{\al \alp}{r^{2}}   \Om^2 \di u \di \ub \nnb \\
	& +\iint_{\D} 2 \(  2 \(  w  r \al +  w^2 \al^2  \) - 4w \Ph  \al - 7 r \Ph  \)  \frac{ \al^2}{r^{2}} \Om^2 \di u \di \ub.
} 
We note that, terms on the left-hand side of \eqref{E:eid-sum2} are all positive. In particular, the energy on $\Cub$ is coercive (the energy on $C_u$ is obviously coercive), and there is
\alis{
	& \int_{\Cub}  \frac{h^{-1}}{r} \( \Ph^2 + \al^2 + h^2 (\p_u w)^2 \)  \di u \\
	\lesssim &  \int_{\Cub}  \frac{h^{-1}}{r} \( 2 \(   w \al - \Ph \)^2 + \frac{ \Ph^2 }{2} + 2 h^2 (\p_u w)^2 \)  \di u.
}
In view of the bootstrap assumption \eqref{BT-H1}, the smallness of  $\mu$, and the boundedness of $r$ in $\D$, the spacetime integrals on the right-hand side of \eqref{E:eid-sum2} can be absorbed by the bulk integrals on the left-hand side. We thus achieve the energy bound 
\ali{Energy-bound-pf}{
	&\int_{C_u} \frac{\Om^2 h}{r}   \(  \al^2 +  (\alp)^2  +  \Ph^2 \) \di \ub \nnb\\
	&+ \int_{\Cub}  \frac{h^{-1}}{r} \( \Ph^2 +  \al^2 + h^2 (\p_u w)^2 \)  \di u \lesssim I_1.
}
As a result, the energy bound \eqref{E:lem-energy-bound} follows.

\subsection{Pointwise estimates}\label{sec-pt-H1}
With the help of energy estimates derived earlier, we now proceed to establish various $L^\infty$ estimates in order to improve the bootstrap assumption.

\bg{lemma}\label{lem-pw-mu}
We have the enhanced estimates for $\mu$ \[ \lim_{u \rightarrow \ub_\ast} \sup_{\D(u, \, \ub_\ast )} \mu  =0, \quad \text{and} \quad \mu|_\Ga = 0. \] 
\ed{lemma}

\bpf
Integrating \eqref{eq-Lm} in $\ub$, we have the mass formula,
\[ m= \int_{0}^r  \( \(1-\mu \)   \al^2 + \Ph^2 \) \di r'|_{C_u}. \]
Due to the smallness of $\mu$, there is the equivalence 
\beq\label{equi-m-E}
 m \sim  \int_0^r \( \al^2 + \Ph^2 \) \di r^\prime|_{C_u}. 
\eeq
Since $0\leq r'\leq r$, and hence $\frac{1}{r} \leq \frac{1}{r'}$, the energy bound \eqref{E:Energy-bound-pf} leads to
\alis{
\frac{m}{r } \leq \int_{0}^r \frac{ \( \(1-\mu\)   \al^2 + \Ph^2 \) }{r^{\prime}} \di r'|_{C_u} \leq E_0 \lesssim I_0.
}
This lemma is concluded. 

\epf

As a next step, we will improve the pointwise estimates for $\al$ and $r^{-1} \ha$.

\bg{lemma}\label{lem-pw-al-Ph}
The estimates for $\al$ and $r^{-1} \ha$ are enhanced as follows
\[r^{-2} \ha^2 \lesssim  \vep^2 \quad \text{and} \quad \al^2 \lesssim \vep, \] and
\[r^{-2} \ha^2|_\Ga = 0  \quad \text{and} \quad \al^2|_\Ga =0.	\]
 
\ed{lemma}

\bpf[Proof of Lemma \ref{lem-pw-al-Ph}]
Multiplying with $2\al$ on \eqref{eq-pu-al} yields \[ \p_u \al^2 =-2\frac{h^{-1}}{r} w \Ph \al+ 2 \frac{h^{-1}}{r} \(  \mu -  \Ph^2  \) \al^2. \]
Then integrate in $u$ on each $\Cub$,
\alis{
  \big| \al^2 - \al^2 |_{ C_{u_0} }  \big| 
\lesssim {}& \int_{\Cub}  h^{-1} \frac{\Ph^2}{r} \di u + \int_{\Cub} h^{-1} w^2 \frac{\al^2}{r} \di u \\
& + \int_{\Cub} 2h^{-1} \( \mu + \Ph^2 \) \frac{\al^2}{r} \di u \\
\lesssim {}& \Eb_0^\prime \lesssim I_1,
}
where the bootstrap assumption \eqref{BT-H1} and the energy bound \eqref{E:Energy-bound-pf} are used.
That is, we have derived a preliminary estimate for $\al$, \[ |\al|^2 \lesssim \al^2 |_{ C_{u_0}}  + I_1. \]  

Based on the boundedness of $\al$, we can further refine its estimate. Noting that the boundedness of $\al$ tells $r \al^2|_{\Ga} =0$, we integrate $\p_r (r \al^2)$ in $r$ on each $C_u$ to yield 
\alis{
r \al^2 = {}& \int_0^r \( \al^2 + 2\al \alp \) \di r'|_{C_u} \\
\leq {}&  \int_0^r  \al^2 \di r'|_{C_u} + \( \int_0^r  \al^2 \di  r'|_{C_u}  \)^{\frac{1}{2} } \( \int_{0}^r (\alp)^2 \di  r'|_{C_u}  \)^{\frac{1}{2}} \\
\lesssim {}& m + m^{\frac{1}{2}} r^{\frac{1}{2}} I_1^{\frac{1}{2}}.
}
Here in the last inequality,  we have access to  the equivalence \eqref{equi-m-E} and the energy bound \eqref{E:Energy-bound-pf}. As a result, 
\[ \al^2 \lesssim \mu+ \mu^{\frac{1}{2}}. \]
Combined with Lemma \ref{lem-pw-mu}, the first part of Lemma \ref{lem-pw-al-Ph} is justified.

The estimate for $r^{-1}\ha$ (or $\Ph$) follows in an analogous fashion.
 We will first show that $r^{-1} \ha$ is bounded by means of integrating $\p_u \( r^{-2} \ha^2\)$ in $u$ on each $\Cub$,
 \alis{
  \frac{\ha^2}{r^2}\Big|_{u_0}^u = {}& \int_{\Cub} \(  \p_u w \frac{2 \ha}{r^2} - 2 \hb \frac{\ha^2}{r^3} \) \di u \\
\lesssim {}&  \int_{\Cub} \frac{h^{-1}}{r} \(  h^2(\p_u w)^2  + \frac{\ha^2}{r^2} \) \di u \lesssim \Eb_0 \lesssim I_0.
}
Therefore, $r^{-2} \ha^2$ is bounded as \[ r^{-2} \ha^2 \lesssim r^{-2} \ha^2\Big|_{C_{u_0}} +I_0, \] which further entails $r \cdot \(r^{-2} \ha^2\)|_{\Ga} =0$. It enables us to integrate  $\p_r\(r \cdot \(r^{-2} \ha^2 \) \)$ in $r$ on each $C_u$ and then an application of the Cauchy-Schwarz inequality leads to
 \alis{
r  \frac{\ha^2}{r^2} = {}& \int_0^r \(  \frac{2 \al \ha}{r} -  \frac{\ha^2}{r^2} \) \di r'|_{C_u} \\
\lesssim {}&  \int_0^r \( 2 \al^2 +  r^{-2} \ha^2 \) \di r'|_{C_u} \lesssim m.  
}
We arrive at
\alis{
 r^{-2} \ha^2 \lesssim \mu,
}
and thus obtain the concluding estimate for $r^{-1} \ha$ with the help of Lemma \ref{lem-pw-mu}.
\epf

With Lemma \ref{lem-pw-al-Ph} concluded, we have improved the pointwise estimates for $r^{-1} \ha$ and $\al$, thereby closing the bootstrap argument. In the subsequent analysis, we will derive more precise estimates to provide a detailed description for the solution.

\subsection{Estimates for the geometry}\label{sec-esti-geometry}

\begin{lemma}
Given the assumptions in Lemma \ref{lem-extension}, the following estimates hold in $\D$,
\[
 |\ln (\Om^2 h)| + | \ln (-\hb) | + |\ln h| \lesssim I_0. 
\]
\end{lemma}

\bpf
Before going into the pointwise estimates, we  deduce some integral bounds for $\mu$.

Denote $r_0 = r(u_0, \, \ub)$. Based on \eqref{eq-Lbmu}, an integration in $u$ along each $\Cub$ yields
\ali{ibp-Lbm}{
& \mu|_{r}^{r_0} + \int_{r}^{r_0} \frac{\mu}{r^{\prime}} \di r^\prime|_{\Cub} \nnb \\ 
={}& \int_{\Cub}  \(2 h^2 \frac{ ( \p_{u} w)^2 }{r} + \( 1-  \mu \) \frac{\Ph^2}{r} \) h^{-1} \di u \lesssim I_0,
} 
where the energy bound \eqref{E:lem-energy-bound} is employed in the last inequality.
Since $ \mu \leq \vep^2$, we thus obtain
\beq\label{bound-mu-H1-Cub}
  \int_{\Cub} (1 - \mu) \frac{\mu}{r} h^{-1}\di u = \int_{r}^{r_0} \frac{\mu}{r^\prime} \di r^\prime \big|_{\Cub} \lesssim I_0 + \vep^2 
\eeq
along each incoming cone $\Cub \cap \D$, $\ub \in (u_0, \, \ub_\ast + \zeta]$. 

In the same manner, due to \eqref{eq-Lmu}, there is, along each $C_u \cap \D$, $u \in [u_0, \, u_1]$, 
\ali{ibp-Lm}{
 & \mu|_{0}^r + \int_{0}^r \frac{\mu}{r^{\prime}} \di r^\prime|_{C_u} \nnb \\ 
={}& \int_{0}^r  \(2  \(1-  \mu \) \frac{\al^2}{r^\prime} + \frac{\Ph^2}{r^\prime} \) \di r^\prime|_{C_u} \lesssim I_0,
}
which leads to
\beq\label{bound-mu-H1-Cu}
   \int_{C_u} \frac{\mu}{r} \Om^2 h \, \di \ub = \int_{0}^r \frac{\mu}{r^\prime} \di r^\prime\big|_{C_u} \lesssim I_0 + \vep^2
\eeq
   along each outgoing cone $C_u \cap \D$, $u \in [u_0, \, u_1]$. 

Now to prove this lemma, let us start with $\ln (\Om^2 h)$. We rewrite the equation \eqref{eq-L-Omh} alternatively as
\[ \p_u  \ln (\Om^2 h) =  - h^{-1} \( \frac{\mu}{r} - \frac{\Ph^2}{r}\). \]
Using the gauge choice \eqref{gauge-ub}, the integral bound \eqref{bound-mu-H1-Cub} and the energy bound \eqref{E:lem-energy-bound}, we derive via an integration in $u$ that,
\[ \Big| \ln \( 2\Om^2 h  \)   \Big|  \lesssim  \int^u_{u_0} \( \frac{\mu}{r} + \frac{\Ph^2}{r} \)  h^{-1}  \di u \lesssim  I_0. \]

After the estimate for $\ln (\Om^2 h)$, we turn to $\ln (-\hb)$. The equation  \eqref{eq-L-hb} indicates \[ \p_{\ub} \( \ln(-\hb) \) =  \frac{\Om^2 h}{1-\mu}  \(\frac{\mu}{r} - \frac{\Ph^2}{r} \) \] which together with the energy bound \eqref{E:lem-energy-bound} and \eqref{bound-mu-H1-Cu} gives rise to \[ | \ln(- \hb)|_{\Ga} - \ln(- \hb) | \lesssim \int_{0}^r  \(\frac{\mu}{r^\prime} + \frac{\Ph^2}{r^\prime} \)  \di r^\prime|_{C_u} \lesssim I_0 + \vep^2. \] Noting the value of $\ln(- \hb)|_{\Ga}$ \eqref{Om2h-hb-Ga}, we then obtain the bound for $ | \ln (-\hb) |$. 

In the end, for $\ln h$, the equation \eqref{eq-Lh} suggests  \[ \p_{\ub} \ln h = - 2\Om^2 h \frac{\al^2}{r}, \] and as a consequence of the energy bound, there is \eqref{E:lem-energy-bound},  \[ | \ln h|_{\Ga} - \ln h| \lesssim \int_{0}^r   \frac{\al^2}{r^\prime} \di r^\prime|_{C_u} \lesssim I_0^2. \] In view of Lemma \ref{lem-pw-mu}, $1+h \hb|_{\Ga} = \mu|_{\Ga}  =0$, and hence $h|_{\Ga} = - \hb^{-1}|_{\Ga} = (\Om^2 h)^{-1}|_{\Ga}$, the estimate for $\ln h$ is proved. 
\epf

\subsection{Proof of Theorem \ref{extension thm-H1}}
For the SSEYM system, the first singularity must emanate from the axis. For further details, we refer to Theorem \ref{extension-thm-away}. It therefore suffices to focus on a small neighbourhood of the tip $u = \ub =\ub_\ast$. We may assume $u_0$ is close to $\ub_\ast$ and that \[ \mu \leq \vep, \quad \text{on} \, \, \D(u_0; \, \ub_\ast) \] 
As indicated by Lemma \ref{lem-extension}, the solution ($\al$) extends as an $H^1$ function to the closure of domain $\D(u_0; \, \ub_\ast)$.  To complete the proof Theorem \ref{extension thm-H1}, it remains to show that the solution can be extended beyond $\Cb_{\ub_\ast}$. Specifically, we must demonstrate that the solution can be extended to $\D(u_0, \, \ub_\ast; \, \ub_\ast + \zeta)$ for some sufficiently small $\zeta>0$. 

We assume that   
\begin{equation}\label{bound-mu-extend}
 \mu \leq \frac{\vep_1 + \vep}{2} \,\, \text{in} \, \, \D(u_0, \, u_1; \, \ub_\ast+\zeta), \quad \text{for any} \,\, u_1 \in (u_0, \, \ub_\ast], 
\end{equation} 
where $\vep_1 > \vep$ is a small constant.

Define \[ u_2 = \sup \,\left\lbrace u_1 \in (u_0, \, \ub_\ast]: \,\,  \sup_{ \D (u_0, \, u_1; \, \ub_\ast+\zeta) } \mu \leq  \frac{\vep_1 + \vep}{2} \right\rbrace. \] Then either $u_2 = \ub_\ast$ or $u_2 < \ub_\ast$ and $\sup_{ \D (u_0, \, u_2; \, \ub_\ast+\zeta) } \mu = \frac{\vep_1 + \vep}{2}$.
Now we shall exclude the second possibility. This will be proved by showing that the bootstrap assumption \eqref{bound-mu-extend} can be improved as \[ \mu < \frac{\vep_1 + \vep}{2}, \quad \text{in} \, \, \D(u_0, \, u_1; \, \ub_\ast+\zeta), \quad \text{for any} \,\, u_1 \in (u_0, \, \ub_\ast], \] if $\zeta$ is small enough.

Preliminarily, by virtue of the local existence theorem (Section \ref{sec-local}) and Lemma \ref{lem-extension}, the bootstrap assumption \eqref{bound-mu-extend} leads to an $H^1$ solution in $\D(u_0, \, \ub_\ast; \, \ub_\ast+\zeta)$. In particular, the $H^1$ norm 
\beq\label{H1-D}
 \int_{C_u^{[\ub_\ast, \,\ub_\ast + \zeta ]} } \frac{\Om^2 h}{r} \( \Ph^2 + \al^2 + (\alp)^2 \) \di \ub, \quad u \in [u_0, \, u_1],
\eeq is uniformly bounded.

On any $C_u$, $u\in [u_0, \, \ub_\ast]$,
an integration in $\ub^\prime \in [\ub_\ast, \, \ub]$ as \eqref{E:ibp-Lm} yields the identity
\alis{
\mu \Big|_{\ub_\ast}^{\ub} +  \int_{r(u, \, \ub_\ast)}^{r(u, \, \ub )} \frac{\mu}{r} \di r|_{C_u} =   \int_{\ub_\ast}^{\ub} \frac{\Om^2 h}{r} \( 2 \(1- \mu \)   \al^2 + \Ph^2 \) \di \ub^\prime,
}
where $\ub \in [ \ub_\ast, \, \ub_\ast + \zeta]$.
Then $\mu$ is bounded as \[  \mu (u, \, \ub )  <  \mu (u, \, \ub_\ast)  +    \int_{\ub_\ast}^{\ub} \frac{\Om^2 h}{r} \( 2 \al^2 + \Ph^2 \) \di \ub^\prime, \quad \ub \in [\ub_\ast, \, \ub_\ast + \zeta]. \]
Due to the uniform boundedness of the $H^1$ norm \eqref{H1-D}, 
we can take $\zeta$ small enough, such that \[  \int_{\ub_\ast}^{\ub} \frac{\Om^2 h}{r} \( \al^2 + \Ph^2 \) \di \ub^\prime < \frac{\vep_1 - \vep}{2}, \quad  \ub \in [\ub_\ast, \, \ub_\ast + \zeta]. \]  Thus, making use of the bootstrap assumption \eqref{bound-mu-extend} improves the upper bound for $\mu$ \[ \mu (u, \, \ub)  <  \mu (u, \, \ub_\ast)  + \frac{\vep_1 - \vep}{2} < \frac{\vep_1 + \vep}{2}.  \] 

In summary, we have proved \eqref{bound-mu-extend} holds with $u_1 = \ub_\ast$. Consequently, following the argument leading to Lemma \ref{lem-extension}, the solution extends as an $H^1$ solution to the domian $\D(u_0, \, \ub_\ast; \, \ub_\ast + \zeta)$. Together with the local existence theorem (Section \ref{sec-local}) for $H^1$ data, it allows us to further extend the solution to $\D(\ub_\ast; \, \ub_\ast + \zeta^\prime)$ for some $\zeta^\prime \in (0, \, \zeta]$. This completes the proof of Theorem \ref{extension thm-H1}.

\subsection{Cauchy stability in $H^1$ space}\label{sec-CS}

Having established the extension theorem (Theorem \ref{extension thm-H1}), we now turn to Cauchy stability within the same topology. This amounts to proving a small-data global existence on the finite domain $\D(u_0, \, \ub_\ast; \ub_\ast)$, where $\ub_\ast \in (u_0, \, + \infty)$. Recall that the energies $E_i$, $i=0, 1$, and $\Eb_0$, $\Eb_0^\prime$, are defined as in \eqref{E:def-E-Eb}. We now state the Cauchy stability theorem.

\begin{theorem}\label{stability-thm-1}
Suppose on the initial hypersurface $C_{u_0}$ the initial energy is small
\[ E_1 (u_0, \, \ub_\ast)  < \vep^2, \] with $\vep$ sufficiently small.
Then there exists a unique and global solution $\( r, m , \al \) \in C^1 \times C^0 \times H^1$ in the domian $\D(u_0,\, \ub_\ast; \ub_\ast)$  such that  \[  r^{-2} |\ha|^2  + |\al|^2 + \mu  \lesssim \vep^2, \] and for $u_0 <u \leq \ub_\ast$, $u_0 < \ub \leq \ub_\ast$, the solution satisfies the following unifom energy bound
\beq\label{thm-energy-small}
E_1 (u, \, \ub) + \Eb^\prime_0 (u, \, \ub) \lesssim \vep^2.
\eeq      
\end{theorem}

\bpf
The proof goes through in line with the one leading to Theorem \ref{extension thm-H1}, and we therefore omit the details here.

 \epf

\section{Local existence theorem}\label{sec-local}

 For the characteristic initial value problem with data prescribed on a future light cone emanating from the center of symmetry, the local existence theorem for classical solutions to the spherically symmetric Einstein scalar field system was proved in \cite{C-global-86}. Regarding to the proof, Christodoulou reduced the whole system in Bondi coordinates to a scalar equation and utilized the contracting mapping principle to obtain the existence theorem. 
 
In this section, however, we aim to present a proof based on energy methods. In more details, we work in the double null gauge and demonstrate a local existence theory for the SSEYM system in $H^1$, using an iteration method.

The Yang--Mills equation can be alternatively written in terms of $\ha$  \eqref{def-ahat}
\beq\label{YM-1}
\p_u \p_{\ub} \ha +2  \frac{\Om^2}{r^2} \ha =  \Om^2 \frac{ \ha^2}{r^2} \(\ha+3\).
\eeq

\begin{theorem}\label{thm-local-H2}
Let $\ub_1 \in (u_0, \,+\infty)$ be a finite constant.
Suppose on the initial slice $C_{u_0}$, the energy is bounded \[  \int_{C_{u_0} } r^{-1} \( \frac{\ha^2}{r^2}  + \al^2 + (\alp)^2 \) \di r <I_1. \]

i) There exists $u_\ast > u_0$ such that the SSEYM system admits a unique solution $\(r, \, m, \, \al \) \in C^1 \times C^0 \times H^1$ in the domain $\D(u_0, u_\ast; \ub_1)$, satisfying the bound,
\alis{
& \int_{C_u} r^{-1} \( \frac{\ha^2}{r^2}  + \al^2 + (\alp)^2 \)  \di  r  \lesssim I_1, \\
& \int_{\Cub}  \frac{h^{-1}}{r } \(  \frac{\ha^2}{r^2} +  \al^2 + h^2 (\p_u \ha)^2 \)  \di u \lesssim I_1,  \\
& | r^{-1} \ha |^2 + |\al|^2 + \mu \lesssim I_1,\\
& | \ln\(\Om^2h\) |, \, |\ln\(-\hb\)|, \, |\ln h| \lesssim I_1.
}

ii) The solution depends continuously on the initial data in the following sense.  Suppose $\(r_n, \, m_n, \, \ha_{n} \)$ and $\(r_m, \, m_m, \, \ha_{m} \)$ are solutions with data $\ha_{0n}$, $\ha_{0m}$ respectively, then
 \begin{align*}
& \int_{C_u} \(  \frac{|\ha_{n} - \ha_m|^2 }{ r^{3}_{n}} +  \frac{|\al_{n} - \al_m|^2 }{ r_{n}} + \frac{|\alp_{n} - \alp_m|^2 }{ r_{n}} \) \di r_n  \\
&+  \int_{\Cub}  \frac{h_n^{-1}}{r_n} \(  \frac{\( \ha_n - \ha_m \)^2}{r_n^2} + \( \al_n - \al_m \)^2 + h_n^2 (\p_u \ha_n - \p_u \ha_m)^2 \)  \di u \\
 \lesssim {}&  \int_{C_{u_0}} \(  \frac{|\ha_{0n} - \ha_{0m}|^2 }{ r^{3}_{n}} +  \frac{|\al_{0n} - \al_{0m}|^2 }{ r_{n}} + \frac{|\alp_{0n} - \alp_{0m}|^2 }{ r_{n}} \) \di r_n .
\end{align*}
\end{theorem}

\bpf

It suffices to prove the local existence theorem near the axis, which essentially reduces the problem to a small-data regime. Away from the axis, local existence follows in a standard way. 
 
Choose $\ub_1 > u_0$ such that $\ub_1 - u_0$ is sufficiently small, ensuring that the initial energy remains sufficiently small.  We denote the size of the initial data by $\vep^2 I_1$, where $\vep^2$ is small enough, namely,  \[ \int_{C_{u_0}^{[u_0, \, \ub_1]} } r^{-1} \( \frac{\ha^2}{r^2}  + \al^2 + (\alp)^2 \) \di r \leq \vep^2 I_1. \]

We shall find a local solution for the SSEYM system using the method of Picard iteration.

{\bf Linearized system.}
We let  $\( r_0, \, \ha_0, \, m_0 \) = \( r|_{C_{u_0}} + \frac{u_0-u}{2}, \, 0, \, 0 \)$ and then define $\(r_{k+1}, \, m_{k+1}, \, \ha_{k+1} \)$ iteratively for $k \geq 0$ as the unique solution to the linearized system:
\begin{align}
 \p_u \p_{\ub} \ha_{k+1} +{}& 2 \Om_k^2 \frac{\ha_{k+1} }{r_{k}^2} =   \Om_k^2 \frac{ \ha_k^2}{r_k^2} \( \ha_k+3 \), \label{eq-L-YM} \\
 \p_{\ub} m_{k+1} ={}& \Om^2_k h_k \( \frac{\Ph_{k+1}^2}{2} - h_k \hb_k  \al_{k+1}^2\), \label{eq-L-Lm} \\
 \p_u m_{k+1} ={}& \hb_k \( \frac{\Ph_{k+1}^2}{2} - \(h_k \hb_k\)^{-1}  h^2_k (\p_u \ha_{k+1})^2 \), \label{eq-L-Lbm} \\
\p_u \(  \Om^2_{k+1} h_{k+1} \) = {}&  \Om^2_{k+1}  \hb^{-1}_{k+1} \hb_k \( - \frac{2m_{k+1}}{r^2_k} +  \frac{ \Ph_{k+1}^2 }{r_k} \),  \label{eq-L-LbLr} \\
\p_{\ub} \(  \hb_{k+1} \) ={}&  \Om^2_k h_k h_{k+1}^{-1}  \( - \frac{2m_{k+1}}{r^2_k} +  \frac{  \Ph_{k+1}^2}{r_k} \),   \label{eq-L-Lhb}
\end{align}
with initial data prescribed in Theorem \ref{thm-local-H2} and subject to the boundary conditions  \[ m_{k+1}|_{\Ga} =0, \quad r_{k+1}|_{\Ga} =0, \quad \ha_{k+1} |_{\Ga} =0. \] 
Here we set
\alis{
 \Om_{k+1}^2 h_{k+1} :={}& \p_{\ub} r_{k+1}, &\hb_{k+1} := \p_u r_{k+1}, \\
\Ph_{k+1} :={}&  \frac{ ( w_{k+1}^2-1 )}{r_k}, & \al_{k+1} := \Om^{-2}_k h_k^{-1} \p_{\ub} \ha_{k+1}, 
 }
and $h_{k+1}$ is related to $m_{k+1}$ via
\[ 2m_{k+1} := r_{k} \( 1+ h_{k+1} \hb_{k+1} \). \] We further define \[ \mu_{k+1} := \frac{2m_{k+1}}{r_k} =\( 1+ h_{k+1} \hb_{k+1} \), \quad  \alp_{k+1} := r_k \Om^{-2}_k h_k^{-1} \p_{\ub} \al_{k+1}. \] 

We insist the choice of gauge in Section \ref{sec-gauge} so that 
\beq\label{gauge-app}
\Om_{k + 1}^2 h_{k + 1} |_{C_{u_0}} =\frac{1}{2}.
\eeq
 Besides, our gauge choice \[ u=\ub \quad  \text{on} \,\, \Ga \] together with the boundary conditions $r_{k+1}|_{\Ga} =0$,  $\ha_{k+1}|_{\Ga} =0$ entails 
\alis{
 \( \hb_{k+1} + \Om^2_{k+1} h_{k+1} \) |_{\Ga} =0, \\
  \( \p_u \ha_{k+1} + \p_{\ub} \ha_{k+1} \) |_{\Ga} =0.
 }    
We remark that, for the first generation, we have $\hb_0=-\frac{1}{2}$, and $h_0=2$, $\Om^2_0=\frac{1}{4}$. In other words, the first generation is the vacuum spacetime.

In the linearized system \eqref{eq-L-YM}-\eqref{eq-L-Lhb}, we first solve \eqref{eq-L-YM} to determine $\ha_{k+1}$. Substituting the $\ha_{k+1}$ into \eqref{eq-L-Lm} and \eqref{eq-L-Lbm}, we then solve $m_{k+1}$. Finally, substituting both of $\ha_{k+1}$ and $m_{k+1}$ into \eqref{eq-L-LbLr} and \eqref{eq-L-Lhb}, we are able to solve $\Om^2_{k+1} h_{k+1}$ and $\hb_{k+1}$, thereby obtaining $r_{k+1}$.


{\bf Auxiliary equations.}
Along with the main equations, we can derive the following two equations from \eqref{eq-L-Lm} and \eqref{eq-L-Lhb}, \eqref{eq-L-Lbm} and \eqref{eq-L-LbLr}.
\begin{align}
\p_{\ub} h_{k+1} 
={}& 2 \Om^{2}_k h_{k} (1-\mu_k) \hb_{k+1}^{-1}  \frac{( \al_{k+1})^2}{r_k},  \label{eq-L-Lh} \\
\p_u (\Om^{-2}_{k+1} \hb_{k+1}) = {}& - 2 \Om^{-2}_{k+1} h^{-1}_{k+1} h_k \frac{(\p_u \ha_{k+1})^2}{r_k}. \label{eq-L-Lbhb}
\end{align}

With the above setup, we now continue the proof with $H^1$ estimates.
 
{\bf Uniform bound.}  We assert that if $\vep$ (or equivalently, $\ub_1-u_0$) is small enough, there is a constant $M<+\infty$ such that the uniform bounds, 
\ali{Uniform-energy-k-local}{
A_{i+1}(u, \ub) :=& \int_{C_u} \Om^2_i h_i \(  \frac{( \alp_{i+1})^2}{ r_i} +  \frac{\al_{i+1}^2}{ r_i} +  \frac{\ha_{i+1}^2}{ r^3_i}  \)  \di u \nnb \\
& +  \int_{\Cub} h_i^{-1} \( \frac{\al_{i+1} ^2 }{r_{i}} +h_i^2  \frac{(\p_u \ha_{i+1})^2}{r_{i}}  +\frac{\ha_{i+1}^2}{r^3_{i}}  \)  \di u \nnb \\
  \leq {}& \vep^2 M,  \quad u \in [u_0, \, u_\ast], \,\, \ub \in [u, \, \ub_1],
}
hold for all $i \geq 0$,

When $i=0$, an energy argument (in line with the proof leading to \eqref{E:eid-sum2} in the subsection \ref{subsec-energ-est-H2}) for the linearized equation  \[\p_u \p_{\ub} \ha_{1} + 2 \Om^2_0 \frac{\ha_{1} }{r_{0}^2} = 0, \] yields
 \begin{align*}
& \int_{C_u} \frac{\Om^2_0 h_0}{r_0}  \(  ( \alp_{1})^2 + 9 \al_{1}^2 + 4  r_0^{-2} \ha_{1}^2  \)  \di \ub  \\
& +  \int_{\Cub} \frac{h_0^{-1}}{r_{0}  } \( 2  \al_{1} ^2 + 2 h_0^2 (\p_u \ha_{1})^2 + 8 r_0^{-2} \ha_{1}^2 \)  \di u   \\
& + \iint_{\D} \frac{ \Om^2_0}{ r^{2}_0} \( 18 r^{-2}_0 \ha_{1}^2 + 2 h_0^2 (\p_u \ha_{1})^2 +\al^2_{1} +   (\alp_{1})^2 \) \di u \di \ub   \\
 ={}& \int_{C_{u_0}}  \frac{\Om^2_0 h_0}{r_0}  \(  ( \alp_{1})^2 + 9 \al_{1}^2 + 4  r_0^{-2} \ha_{1}^2  \)  \di \ub. 
\end{align*}
Thus, \eqref{E:Uniform-energy-k-local} holds for $i=0$, which further implies
\alis{
& \mu_1 + \al^2_{1} + \frac{\ha^2_{1}}{r_{0}^2} \lesssim \vep^2 I_1, \\
& |\ln (\Om_1 h_1)|, \,  |\ln (-\hb_1)|, \, |\ln h_1| \lesssim I_1.
  } 

To proceed,  let us suppose the estimate \eqref{E:Uniform-energy-k-local} holds for $i \leq k-1$ with the constant $M$ to be determined later. 
Then we can derive the pointwise estimates, as in the subsections \ref{sec-pt-H1} and \ref{sec-esti-geometry}
\alis{
& \mu_{i+1} + \al^2_{i+1} + \frac{\ha^2_{i+1}}{r_{i}^2} \lesssim \vep^2 M, \quad i \leq k-1, \\
& |\ln (\Om_{i+1} h_{i+1})|, \,  | \ln (-\hb_{i+1})|, \, |\ln h_{i+1}| \lesssim I_1, \quad  i \leq k-1.
  } 
 Moreover, using the equation of \eqref{eq-L-LbLr} with $k$ replaced by $k-1$ or $k-2$, and noting the gauge choice \eqref{gauge-app}, we have \[ \Big| \ln \frac{ \Om_{k-1}^2 h_{k-1} }{ \Om_{k}^2 h_{k} } \Big| \lesssim \vep^2 M, \quad \Big| \ln \frac{ \Om_{k-2}^2 h_{k-2} }{ \Om_{k-1}^2 h_{k-1}  } \Big|  \lesssim \vep^2 M. \] Here we set $r_{i} = r_0$, $\Om^2_i h_i = \Om^2_0 h_0$ if $i<0$. These estimates plus the identity \[ \Om_{k-1}^{-2} h^{-1}_{k-1} \p_{\ub} (r_{k} - r_{k-1}) = \(-1 + \frac{\Om_{k}^2 h_{k}}{\Om_{k-1}^2 h_{k-1}} \), \] suggest the equivalence of $r_k$, $r_{k-1}$, $r_{k-2}$, if $\vep$ is small enough,
 \[ \frac{1}{2} < \big| \frac{r_{k}}{r_{k-1}}   \big|, \,\,  \big| \frac{r_{k-1}}{r_{k-2}}   \big| < \frac{3}{2}. \] 
 In addition,  taking the equation of \eqref{eq-L-Lhb} (with $k$ replaced by $k-1$) and the gauge condition of $\hb_{k} |_{\Ga} = - \Om^2_{k} h_{k} |_{\Ga}$ into account, we find that  \[ \Big| \ln \frac{ \hb_{k-1} }{ \hb_{k} } \Big| \lesssim \vep^2 M, \] and hence \[ \Big|  \ln \frac{ h_{k-1} }{ h_{k} } \Big| \lesssim \vep^2 M \] follows, since $\frac{ h_{k-1} }{ h_{k} } = \frac{ h_{k-1} \hb_{k-1}}{ h_{k} \hb_{k}} \cdot \frac{ \hb_{k} }{ \hb_{k-1} }$.
 
 We shall then prove  \eqref{E:Uniform-energy-k-local} holds for $i=k$. 
Thanks to the preceding estimates and an energy argument analogous to that in Subsection \ref{subsec-energ-est-H2}, we are able to derive
 \begin{align}
& \int_{C_u} \Om^2_k h_k \(  \frac{( \alp_{k+1})^2}{ r_k} + 9 \frac{\al_{k+1}^2}{ r_k} + 4 \frac{\ha_{k+1}^2}{ r_k}  \)  \di \ub \nnb \\
& +  \int_{\Cub} \frac{h_k^{-1}}{r_{k}}\( 2 \(  \al_{k+1} -2 r_k^{-1} \ha_{k+1} \)^2 + 2 h_k^2 (\p_u \ha_{k+1})^2 + 2 r_k^{-2} \ha_{k+1}^2 \)  \di u \nnb \\
& + \iint_{\D} \frac{1}{2} \Om^2_k r^{-2}_k \( 18 r^{-2}_k \ha_{k+1}^2 + 2 h_k^2 (\p_u \ha_{k+1})^2 + \al^2_{k+1} +   (\alp_{k+1})^2 \) \di u \di \ub \nnb \\
 \leq {}& \vep^2 I_1 \cdot C, \label{eid-Linear-halp-3}
\end{align}
where the universal constant $C$ is independent of $k$.
If we choose $M > C I_1$, then \eqref{E:Uniform-energy-k-local} with $i=k$ is verified.

 {\bf Convergence.}
  In the compact domain concerned, the sequence $\{ r_k \}$ is equicontinuous. Therefore, there is a subsequence (still denoted by $ \{ r_k \}$) that converges uniformly to a function $r$, $r_k \rightarrow r$, on the compact neighbourhood. In the subsequent analysis, we will always work with this convergent subsequence in place of the original sequence.

Based on \eqref{eq-L-YM}, we derive the following linear equation,
\ali{eq-L-diff}{
 & \p_u \p_{\ub} ( \ha_{k+1} - \ha_{k} ) + 2 \frac{ \Om_{k-1}^2 }{r_{k-1}^2} ( \ha_{k+1} - \ha_{k})  \nnb \\
 = {}& f \(\frac{\ha_{k+1}}{ r_{k}} , \, \frac{\ha_k}{r_{k} }, \, \frac{\ha_{k-1}}{r_{k-1} } \) \cdot  \frac{(U_k - U_{k-1})}{r_k}, 
 }
  where $U_k - U_{k-1}$ involves \[ \frac{r_k - r_{k-1}}{r_{k-1}}, \,\, h^{-1}_k - h^{-1}_{k-1}, \,\, \Om^2_k h_k - \Om^2_{k-1} h_{k-1}, \,\,  \frac{\ha_{k}}{r_{k-1}} - \frac{\ha_{k-1}}{r_{k-2} }, \] and  \[ \Big| f \(\frac{\ha_{k+1}}{ r_{k}} , \, \frac{\ha_k}{r_{k} }, \, \frac{\ha_{k-1}}{r_{k-1} } \) \Big| \lesssim  \big| \frac{\ha_{k+1}}{ r_{k}}\big| + \big|\frac{\ha_k}{r_{k} } \big| + \big|\frac{\ha_{k-1}}{r_{k-1} } \big|. \] 
In general, we expect $|U_k - U_{k-1}|$ can be bounded by $\| \ha_{k} - \ha_{k-1} \|_{H^1}$.
  

We will prove the convergence by induction. Suppose 
\begin{align}
 & \|U_k - U_{k-1} \|_{L^\infty}  \leq M 2^{-k+1}, \quad \|\mu_k - \mu_{k-1} \|_{L^\infty} \leq M 2^{-k+1}, \label{conve-induction-1} \\
&\Big|  \int_{C_u} \frac{m_{k}}{r_{k-1}^2} \di r_{k-1}|_{C_u} - \int_{C_u} \frac{m_{k-1}}{r_{k-2}^2} \di r_{k-2}|_{C_u} \Big| \leq M 2^{-k+1}, \label{conve-induction-2} \\
&\Big| \int_{\Cub} \frac{m_{k}}{r_{k-1}^2} \di r_{k-1}|_{\Cub} - \int_{\Cub} \frac{m_{k-1}}{r_{k-2}^2} \di r_{k-2}|_{\Cub} \Big| \leq M 2^{-k+1}, \label{conve-induction-3}
 \end{align}
 for a large constant $M$ to be determined. When $k=1$, \eqref{conve-induction-1}-\eqref{conve-induction-3} hold by the uniform boundedness in the previous step. We will prove the same inequalities hold with $k$ replaced by $k+1$.

As is evident from its expression, the linear equation \eqref{E:eq-L-diff} shares the same structure as \eqref{eq-L-YM}. Thus, we can establish an analogous energy identity, taking into account that the initial data now vanish,
 \begin{align*}
& \int_{C_u} h_{k-1} \Om^{2}_{k-1}   \frac{|X_{k+1} - X_k|^2 }{ r_{k-1}}  \di \ub  +  \int_{\Cub} h^{-1}_{k-1} \frac{|Y_{k+1} - Y_k|^2 }{ r_{k-1}}  \di u \nnb \\
& + \iint_{\D} \Om^2_{k-1} r^{-2}_{k-1} |V_{k+1} - V_k|^2 \di u \di \ub \nnb \\
 \lesssim{}& \iint_{\D} \( \big| \frac{\ha_{k+1}}{ r_{k}}\big| + \big|\frac{\ha_k}{r_{k-1} } \big| + \big|\frac{\ha_{k-1}}{r_{k-2} } \big| \) \Om_{k-1}^2 \frac{|U_k-U_{k-1}| |V_{k+1} - V_k|}{r_{k-1} r_{k-2}} \di u \di \ub.
\end{align*}
Here, for simplicity, we denote 
\alis{
X_{k+1} - X_{k} ={}& \(r_{k-1} \p_{r_{k-1}}^2 ( \ha_{k+1} - \ha_k) , \, \p_{r_{k-1}} ( \ha_{k+1} - \ha_k),  \, r_{k-1}^{-1} ( \ha_{k+1} - \ha_k )  \),\\
 Y_{k+1} - Y_{k} = {}& \( \p_{r_{k-1}} ( \ha_{k+1} - \ha_k), \, \p_u  ( \ha_{k+1} - \ha_k),  \, r_{k-1}^{-1} ( \ha_{k+1} - \ha_k )  \),\\
V_{k+1} - V_{k} ={}& \( X_{k+1} - X_{k},  \, \p_u  ( \ha_{k+1} - \ha_k) \). 
 }
Then we estimate the right hand side of the inequality as follows,
 \begin{align*}
 \lesssim{}& \|U_k-U_{k-1}\|_{L^\infty} \( \iint_{\D} \frac{\Om_{k-1}^2}{r^2_{k-2}} \( \big| \frac{\ha_{k+1}}{ r_{k}}\big|^2  + \big|\frac{\ha_k}{r_{k-1} } \big|^2  + \big|\frac{\ha_{k-1}}{r_{k-2} } \big|^2 \) \di u \di \ub \)^{\frac{1}{2}} \\
 & \qquad \cdot \( \iint_{\D} \Om_{k-1}^2 \frac{|V_{k+1} - V_k|^2}{r^2_{k-1} } \di u \di \ub \)^{\frac{1}{2}} \\
 \lesssim{}& \vep \|U_k-U_{k-1}\|_{L^\infty} \( \iint_{\D} \Om_{k-1}^2 \frac{|V_{k+1} - V_k|^2}{r^2_{k-1} } \di u \di \ub \)^{\frac{1}{2}}. 
\end{align*}
As a result, an application of the Cauchy-Schwarz leads to
\ali{ee-iteration}{
& \int_{C_u} h_{k-1} \Om^{2}_{k-1}   \frac{|X_{k+1} - X_k|^2 }{ r_{k-1}}  \di \ub  +  \int_{\Cub}  h^{-1}_{k-1} \frac{|Y_{k+1} - Y_k|^2 }{ r_{k-1}}  \di u \nnb \\
& + \iint_{\D} \frac{1}{2} \Om^2_{k-1} r^{-2}_{k-1} |V_{k+1} - V_k|^2 \di u \di \ub \nnb \\
 \lesssim{}& \vep \|U_k-U_{k-1}\|^2_{L^\infty}. 
}

For notational simplification, in what follows, we will denote 
\alis{
& \| \ha_{k+1} - \ha_k \|^2_{H^1(C_u)} \\
:={}& \int_{C_u} \( | \p_{r_{k-1} } \( \ha_{k+1} - \ha_k \) |^2 +  \frac{|\ha_{k+1} - \ha_k|^2 }{r_{k-1}^4} \) \Om_{k-1}^2 h_{k-1}  \di \ub, \\
& \| \ha_{k+1} - \ha_k \|^2_{H^2(C_u)} \\
:={}& \int_{C_u}  | r_{k-1} \p^2_{r_{k-1} } \( \ha_{k+1} - \ha_k \) |^2   \Om_{k-1}^2 h_{k-1}  \di \ub + \| \ha_{k+1} - \ha_k \|^2_{H^1(C_u)}.
}

\underline{Convergence of $\{\ha_k\}$ uniformly  in the energy norm.} 
By \eqref{conve-induction-1} and the fact $\vep$ is small, the estimate \eqref{E:ee-iteration} tells that
 \begin{align}\label{conver-energy}
& \|\ha_{k+1} - \ha_{k}\|^2_{H^2} (C_u) +  \int_{\Cub} h^{-1}_{k-1} \frac{|Y_{k+1} - Y_k|^2 }{ r_{k-1}}  \di u \nnb \\
& \qquad + \iint_{\D} \Om^2_{k-1} r^{-2}_{k-1} |V_{k+1} - V_k|^2 \di u \di \ub \nnb \\
  \leq {}&   \vep C M^2 2^{-2k+2}.
\end{align}
Moreover, by virtue of the fact $\frac{|\ha_{k+1} - \ha_k|^2 }{r_{k-1}^2} \lesssim  \| \ha_{k+1} - \ha_k \|^2_{H^1(C_u)}$,  the previous $L^\infty$ estimates and \eqref{conve-induction-1},
it follows that
\ali{conver-ha}{
\Big| \frac{\ha_{k+1}}{r_k} -  \frac{\ha_{k}}{r_{k-1}} \Big| \lesssim {}& \Big| \frac{\ha_{k+1}}{r_k} \Big| \Big| 1-\frac{r_k}{r_{k-1}} \Big| +  \|\ha_{k+1} - \ha_{k}\|_{H^1}  \nnb\\
\leq {}&  \sqrt{\vep} C M 2^{-k+1} < M 2^{-k},
}
if we choose $\vep$ small enough such that $2 \sqrt{\vep} C < \frac{1}{2}$ and  $M > 1$.

\underline{Convergence of $\mu_k$ (uniformly in $C^0$ norm).}
Noting that 
\ali{diff-mu}{
\frac{m_{k+1}}{r_k} - \frac{m_k}{r_{k-1}} 
={}& \frac{m_{k+1} - m_k}{r_k} + \frac{m_k}{r_{k}}  \( 1- \frac{r_k}{r_{k-1}} \),
}   
and using \eqref{eq-L-Lm}, we obtain 
\alis{
 \frac{m_{k+1} - m_k}{r_k} \lesssim {}&   \| \ha_{k+1} \|^2_{H^1}  \( | U_k - U_{k-1}| + | \mu_k - \mu_{k-1} |  \) \\
 & + \( \| \ha_{k+1} \|_{H^1} +  \| \ha_{k} \|_{H^1} \) \|\ha_{k+1} - \ha_k\|_{H^1}.
 }
Then it follows from the inductive assumptions \eqref{conve-induction-1} and \eqref{conver-energy} that 
\beq\label{conver-mu}
   |\mu_{k+1} - \mu_k| \leq \vep C M 2^{-k+1}. 
\eeq
We conclude through choosing $\vep$ and $M$ properly that \[  |\mu_{k+1} - \mu_k| \leq M 2^{-k}.  \]

Furthermore, in view of the formulae \eqref{E:ibp-Lbm} and \eqref{E:ibp-Lm}, it follows from \eqref{conver-energy} and \eqref{conver-mu} that \eqref{conve-induction-2}--\eqref{conve-induction-3} hold with $k$ replaced by $k+1$. 

\underline{Convergence of $\Om^2_k h_k$, $\hb_k$, $h_k$, $m_k$ (uniformly in $C^0$ norm).}
Similarly, using the equations \eqref{eq-L-LbLr} and \eqref{eq-L-Lhb}, we prove that 
\alis{
& \Big| \ln \frac{\Om^2_{k+1} h_{k+1}}{  \Om^2_k h_k} \Big| \leq \vep C M 2^{-k+1} \leq  M 2^{-k}, \\
 & \Big| \ln \frac{ \hb_{k+1}}{\hb_k} \Big|  \leq \vep C M 2^{-k+1}  \leq M 2^{-k},
 } 
  which together with \eqref{conver-mu} implies $|h_{k+1}^{-1} - h^{-1}_k| \leq \vep M 2^{-k}$. The inductive proof is thus complete.
  
Finally, substituting all the previous results into the equation of mass \eqref{eq-L-Lm}, we obtain \[ m_k - m_{k-1} = O(2^{-k}). \]

Integrating $\Om_{k}^{-2} h^{-1}_{k} \p_{\ub} (r_k - r_{k+1}) = \(1- \frac{\Om_{k+1}^2 h_{k+1}}{\Om_{k}^2 h_{k}} \)$, we have \[ |r_{k+1} - r_k| \lesssim r_k \big|1- \frac{\Om_{k+1}^2 h_{k+1}}{\Om_{k}^2 h_{k}} \big|_{C^0}. \] Then \[ \big| \frac{r_{k+1}}{r_k}-1 \big|=O(2^{-k}). \] 

Due to the uniform convergences, there exist $\mu$, $m$ and $\Om^2 h$, $\hb$ such that
\alis{
& \mu_k \rightarrow \mu \quad  \text{uniformly  in} \, C^0 \, \text{norm}, \\
& m_k \rightarrow m \quad  \text{uniformly  in} \, C^0 \, \text{norm}, \\
& \Om_k^2 h_k \rightarrow \Om^2 h \quad \text{uniformly  in} \, C^0 \, \text{norm}, \\
& \hb_k \rightarrow \hb \quad \text{uniformly  in} \, C^0 \, \text{norm}.
}
Meanwhile, we have already known that $r_k \rightarrow r$ on this compact domain, hence \[ \p_u r = \hb, \quad \p_{\ub} r = \Om^2 h. \] The fact that $\mu_k$ is bounded entails that $m_k=O(r_{k-1})$ and $m = O(r)$. We then have for the sake of uniqueness, \[ \mu = \frac{m}{r}. \] 
Besides, the identity $h_{k} = \frac{\mu_{k} -1}{\hb_{k}}$ gives rise to \[ h = \frac{\mu-1}{\hb}. \]

As a consequence, it follows that \[ \mu \in C^0, \,\, \ln \Om^2h \in C^0,  \,\, \ln(- \hb) \in C^0, \,\, \ln h \in C^0, \,\, \text{and} \,\,  r \in C^1.\]  
In addition, \eqref{conver-energy} suggests
\alis{
& \int_{C_u}  \Om^{2} h  r^{-1} \( \frac{ \(\ha_{k+1} - \ha_k\)^2}{r^2} + \( \p_r \(\ha_{k+1} - \ha_k\)\)^2 + \( r \p^2_r \(\ha_{k+1} - \ha_k\) \)^2  \) \di \ub \\
&+ \int_{\Cub} h^{-1} r^{-1}  \( \frac{ \(\ha_{k+1} - \ha_k\)^2}{r^2} + \( \p_r \(\ha_{k+1} - \ha_k\)\)^2 + h^2 \(  \p_u \(\ha_{k+1} - \ha_k\) \)^2  \) \di u \\
 ={}& O(2^{-k}).
}
Hence, there exists $\ha  \in C^0$ such that \[ \ha_k \rightarrow \ha \quad \text{uniformly in} \, H^2 \, \text{norm}, \] 
and \[ \frac{\ha}{r}, \, \p_r \ha  \in C^0. \] 
Setting \[ w:=\ha+1,\] we conclude that $(r, \, m,\, w)$ satisfies the SSEYM equations. 
However, since $\p_u \p_r w$ is not necessarily bounded on the axis, the Yang--Mills equation, and indeed the entire SSEYM system, holds in the sense of distribution.

{\bf Continuous dependence on the data.} Suppose $\ha_{n}$ and $\ha_{m}$ are solutions with data $\ha_{0n}$, $\ha_{0m}$ respectively. The following equation holds
\[ \p_u \p_{\ub} ( \ha_{n} - \ha_{m} ) + 2 \frac{ \Om_{n}^2 }{r_{n}^2} ( \ha_{n} - \ha_{m}) = f\( \frac{\ha_n}{r_n}, \, \frac{\ha_{m}}{r_m} \) \cdot \frac{(U_n - U_{m})}{r_n},  \] where $U_n - U_{m}$ involves $\frac{r_n - r_{m}}{r_n}$, $h^{-1}_n - h^{-1}_{m}$, $\Om^2_n h_n - \Om^2_{m} h_{m}$ and $r^{-1}_n \ha_n - r^{-1}_m \ha_m$, and \[ \big| f\( \frac{\ha_n}{r_n}, \, \frac{\ha_{m}}{r_m} \) \big| \lesssim  \big|\frac{\ha_n}{r_n} \big| + \big| \frac{\ha_{m}}{r_m} \big|. \] 
Following the procedures in proving uniform bounds, we can show that property ii) holds.

In the end, the uniqueness follows in a similar way.

\epf

\appendix
\section{An extension principle away from the axis}\label{sec-extension-away-axis}
The extension principle away from the axis had been established for several systems,  including the spherically symmetric Einstein-Vlasov and Einstein charged scalar field models \cite{Dafermos-local, Kommemi-CMP-13}. As a prerequisite for such an extension theorem, a local existence theorem away from the axis is needed, which was also provided in \cite{Dafermos-local} as well. The proof in \cite{Dafermos-local} was based on the fixed point method and the double null gauge. This approach remains valid for the Einstein equations coupled with various matter fields. In this section, we adapt the methods of \cite{Dafermos-local, Kommemi-CMP-13} to develop a local theory to the SSEYM system away from the axis.

We rewrite the SSEYM system \eqref{eq-Lh}--\eqref{eq-L-omb} and \eqref{YM} in terms of $(r, \Om, w)$ as follows,
\begin{subequations}
\begin{align}
r \p_u \p_{\ub} r ={}& - \Om^2 - \p_u r \p_{\ub} r + \Om^2 r^{-2} Q^2, \label{eq-L-Lb-r} \\
r^2 \p_u \p_{\ub} \ln \Om ={}& \Om^2 + \p_u r \p_{\ub} r - 2 \Om^2 r^{-2} Q^2,  \label{eq-L-Lb-Om} \\
\p_u\( \Om^{-2} \p_u r \) = {} & - 2 \Om^{-2} r^{-2} \(\p_u w \)^2,  \label{eq-Lb2-r}\\
\p_{\ub} \( \Om^{-2} \p_{\ub} r \) = {} & - 2 \Om^{-2} r^{-2} \(\p_{\ub} w \)^2,  \label{eq-L2-r}\\
\p_u \p_{\ub} w +{}& \Om^2 r^{-2} w Q =0, \label{eq-L-Lb-w}
\end{align}                                                                
\end{subequations}
where $Q=w^2-1$. In view of the structures of the system \eqref{eq-L-Lb-r}-\eqref{eq-L-Lb-w}, a local existence theorem away from the axis follows in a standard way \cite{Dafermos-local}.

\begin{theorem}

Let $k\geq 0$ and $\Si := [0, d] \times \{0\} \cup \{0\} \times [0, d]$. On $\Si$, let $r, \Om$ be positive functions such that $r \in C^{k+2}(\Si)$ and $\Om \in C^{k+1}(\Si)$, and let $w \in C^{k+1} (\Si)$. Suppose \eqref{eq-Lb2-r}--\eqref{eq-L2-r} hold initially on $[0,d] \times \{0\}$ and $\{0\} \times [0, d]$, respectively. Define 
\alis{
N_u :={}& \sup_{[0,d] \times \{0\}} \{ |\Om|, \, |\Om^{-1}|, \, |\p_u \Om|, \, |r|, \, |r^{-1}|, \, |\p_u r|, \, |\p_u^2 r|, \, |w|, \, |\p_u w| \}, \\
N_{\ub} :={}& \sup_{ \{0\} \times [0, d] } \{ |\Om|, \, |\Om^{-1}|, \, |\p_{\ub} \Om|, \, |r|, \, |r^{-1}|, \, |\p_{\ub} r|, \, |\p_{\ub}^2 r|, \, |w|, \, |\p_{\ub} w| \},\\
N :={}& \sup \{N_u, \, N_{\ub} \}.
 }
 Then there exists a $\de>0$ depending only on $N$ such that there are functions $(r, \Om, w) \in C^{k+2} \times C^{k+1} \times C^{k+1}$ satisfying equations \eqref{eq-L-Lb-r}--\eqref{eq-L-Lb-w} in $[0, \de^\ast] \times [0, \de^\ast]$, where $\de^\ast = \min\{d, \, \de\}$, with the above prescribed data on $\Si$.
\end{theorem}
   
Building upon the local existence theorem established above, we can prove a generalized extension principle away from the axis (referred to as “strongly tame” in the terminology of \cite{Kommemi-CMP-13}) by adapting the methods of  \cite{Kommemi-CMP-13}. 

\begin{theorem}\label{extension-thm-away} 
Let $\(\mathcal{Q}^+, g_{\mu \nu}, w \)$ be the maximal future development of smooth spherical symmetric initial data for the SSEYM system. For $p \in \ove{\mathcal{Q}^+}$ and $q \in I^-(p) \cap \mathcal{Q}^+ / \{p\}$ such that if $\D = J^+(q) \cap J^-(p) / \{p \} \subset \mathcal{Q}^+$ has finite spacetime volume and there are constant $r_0$ and $R$ such that \[ 0<r_0 \leq r(p') \leq R < \infty, \quad p' \in \D, \] then $p \in \mathcal{Q}^+$.
\end{theorem}

The proof relies on the dominant energy condition and the special structure of the Einstein Yang--Mills system. This structure is reflected in the fact that bounding all variables reduces to estimating the integrals \[ \iint_{\D} r^2 T_{u \ub} \di u \di \ub \quad \text{and} \quad \iint_{\D} \Om^2 T_{AB} g^{AB} \di u \di \ub, \] which in the SSEYM case takes the form \[ \iint_{\D} \Om^2 \frac{Q^2}{r^2} \di u \di \ub. \] This quantity can be bounded straightforwardly through integrating \eqref{eq-L-Lb-r}; for further details, we refer to \cite{Kommemi-CMP-13}.

Theorem \ref{extension-thm-away} shows that the “first singularity” in the maximal future development of spherically symmetric asymptotically flat data for the SSEYM system must emanate from the axis.

\end{document}